\documentclass[12pt]{article}

\usepackage{authblk}
\usepackage{multibib}

\usepackage[superscript]{cite}
\usepackage[mathlines]{lineno}
\usepackage{upgreek} 
\usepackage{tabularx}
\usepackage{times}
\usepackage[utf8]{inputenc}
\usepackage[T1]{fontenc}
\usepackage{caption}
\usepackage{sectsty}
\usepackage{lineno}
\sectionfont{\fontsize{13}{15}\selectfont}
\subsectionfont{\fontsize{12}{15}\selectfont}
\usepackage{nameref}
\usepackage{titlesec}

\usepackage{amsmath}
\usepackage{amssymb}
\usepackage{lineno}
\usepackage{graphicx}
\usepackage{color}
\usepackage{bm}
\usepackage{setspace}

\usepackage[usenames, dvipsnames]{xcolor}
\usepackage[colorlinks=true,citecolor=blue,linkcolor=blue,urlcolor=blue]{hyperref}
\usepackage{abstract}

\makeatletter
\renewcommand{\maketitle}{\bgroup\setlength{\parindent}{0pt}
\begin{flushleft}
  \textbf{\@title}\\

  \@author
\end{flushleft}\egroup
}
\makeatother

\title{\noindent {\bf \large Surface charge density wave in UTe$_2$\\\,}}

\author[1,2]{\normalsize Pablo García Talavera}
\author[1,2]{\normalsize Miguel Águeda Velasco}
\author[3]{\normalsize Makoto Shimizu}
\author[1,2]{\normalsize Beilun Wu}
\author[1,2]{\normalsize Óscar Bou Marqu\'es}
\author[4]{\normalsize Georg Knebel}
\author[4,5]{\normalsize Midori Amano Patino}
\author[4]{\normalsize Gerard Lapertot}
\author[6]{\normalsize \normalsize Jacques Flouquet}
\author[4]{\normalsize Jean Pascal Brison}
\author[6]{\normalsize Dai Aoki}
\author[3]{\normalsize Youichi Yanase}
\author[1,2]{\normalsize Edwin Herrera}
\author[1,2]{\normalsize Isabel Guillam\'on}
\author[1,2]{\normalsize Hermann Suderow}

\affil[1]{\small \it Laboratorio de Bajas Temperaturas y Altos Campos Magn\'eticos, Unidad Asociada UAM-CSIC, Departamento de F\'isica de la Materia Condensada and Instituto Nicol\'as Cabrera, Universidad Aut\'onoma de Madrid, E-28049 Madrid, Spain.}
\affil[2]{\small \it Condensed Matter Physics Center (IFIMAC), Universidad Aut\'onoma de Madrid, E-28049 Madrid, Spain.}
\affil[3]{\small \it Department of Physics, Graduate School of Science, Kyoto University, Kyoto 606-8502, Japan.}
\affil[4]{\small \it University Grenoble Alpes, CEA, IRIG-PHELIQS, F-38000 Grenoble, France.}
\affil[5]{\small \it Univ. Grenoble Alpes, CNRS, Grenoble INP, Institut Néel, 38000 Grenoble, France}
\affil[6]{\small \it Institute for Materials Research, Tohoku University, Ibaraki 311-1313, Japan.}

\date{}

\begin{document} 

\baselineskip24pt


\maketitle 


\noindent{\bf The spatially uniform electronic density characteristic of a metal can become unstable at low temperatures, leading to the formation of charge density waves (CDWs). These CDWs, observed in dichalcogenides, cuprates and pnictides, arise from the interplay between the crystal lattice and the electronic structure, which can facilitate charge ordering. However, CDWs are rarely observed in the presence of Kondo screening and heavy fermion quasiparticles. The heavy fermion topological superconductor candidate UTe$_2$ presents a notable exception, exhibiting a CDW whose origin remains elusive. Here we report high resolution scanning tunneling microscopy (STM) experiments that reveal the primitive wavevectors of the CDW in UTe$_2$. This allows us to identify hot spots in the electronic band structure that are connected to the CDW. Although the corresponding wavevectors have apparently no specific influence on some bulk properties, for example on antiferromagnetic fluctuations, we find that they lead to a spatial modulation of the heavy fermion hybridization pattern. We propose that surface induced modifications in the U 5f electron valence enable a novel form of purely electron-driven charge ordering.
}


\paragraph*{The CDW in UTe$_2$.} The unconventional superconductivity observed in UTe$_2$ exhibits characteristics strongly suggestive of spin-triplet Cooper pairing. The superconducting phase diagram, featuring an exceptionally high critical field and reentrant behavior at extreme magnetic fields\cite{Ran2019, KnebelJPSJ2019,Lewin_2023}, anomalous field dependence of the Knight shift\cite{KinjoPRB2023, MatsumuraJPSJ2023} and multiple transitions between superconducting phases\cite{Braithwaite2019, AokiJPSJ2020, LinNPJQuMat2020,ThomasPRB2021,Wu2024, RosuelPRX2023, SakaiPRL2023, KinjoPRB2023}, poses significant challenges to explanations based on conventional s-wave BCS theory. A relevant question concerns the nature of the normal state parent to the remarkable superconducting properties of UTe$_2$.

Scanning Tunneling Microscopy (STM) studies have revealed the presence of a charge density wave (CDW)\cite{Aishwarya2023,Gu2023,Aishwarya2024,LaFleur2024,gu2025}. Analogous to the cuprates\cite{Wu2011,Chang2016,doi:10.1126/science.1242996}, the CDW is linked to a spatially modulated superconducting phase, termed pair density wave (PDW)\cite{Gu2023,annurev:/content/journals/10.1146/annurev-conmatphys-031119-050711}. However, the underlying mechanism linking the CDW to electronic properties remains poorly understood. Notably, structural studies have shown the absence of detectable lattice distortions characteristic of conventional CDW formation\cite{kengle2024absence,kengle2024absence2}, and macroscopic measurements, such as specific heat, ultrasound attenuation or thermal expansion have failed to identify signatures of a phase transition associated with the CDW\cite{PhysRevB.110.144507,PhysRevB.104.205107}. Moreover, the hitherto observed CDW wavevectors lie outside the first Brillouin zone and have no relationship with features of the bulk band structure\cite{Aishwarya2023,Gu2023,Aishwarya2024,LaFleur2024,PhysRevLett.125.237003,PhysRevB.104.L100409,Butch2022,PhysRevLett.124.076401,PhysRevB.106.L060505,liu2024density,shimizu2025}. Here, we report new high resolution millikelvin STM measurements focusing on high magnetic fields to address as closely as possible the behavior of the normal phase. We demonstrate that the previously identified wavevectors are not primitive wavevectors of the CDW lattice and uncover a strong connection between the surface CDW and the bulk electronic band structure.

\paragraph*{Primitive wavevectors of the CDW of UTe$_2$.} The cleaved surface of UTe$_2$, depicted in grey in Fig.\,\ref{FigureIntro}{\bf a}, corresponds to the (011) plane of the orthorhombic crystal structure (space group \#71, $Immm$). The STM topography image in  Fig.\,\ref{FigureIntro}{\bf b} reveals chains formed by Te(2) atoms, with chains of Te(1) located slightly below. Uranium (U) chains are located between the Te(1) and Te(2) surface chains. The atomic arrangement at the surface defines the surface centered rectangular unit cell, indicated by the solid blue lines in Fig.\,\ref{FigureIntro}{\bf b}. The primitive crystal surface unit cell is represented by dashed blue lines in Fig.\,\ref{FigureIntro}{\bf b}. We applied a magnetic field perpendicular to the surface, along a direction within the superconducting phase diagram at $23.7^{\circ}$ from the {\bf b} axis towards the {\bf c} axis direction, as shown in Fig.\,\ref{FigureIntro}{{\bf c}. This magnetic field direction aligns with the onset of the very high field superconducting phase above 35 T\cite{Ran2019, KnebelJPSJ2019,Lewin_2023}. The bulk Brillouin zone, with standard high symmetry point labels, is shown in black lines in Fig.\,\ref{FigureIntro}{\bf d-f}. The previously reported CDW Bragg peaks are shown in orange, and lie outside of both the bulk and the surface first Brillouin zones (more details in Extended Data Fig.\,\ref{FigBZs} and in Supplementary Information, Section 1). In contrast, the Bragg peaks observed in this study, depicted in yellow, are located within the first Brillouin zones of both the bulk and the surface lattices.

We present atomic resolution tunneling conductance maps acquired near zero bias from 10\,T to 20\,T in Fig.\,\ref{FigureCDW}{\bf a-c}, illustrating data from three different fields of view. The corresponding Fourier transforms are shown in Fig.\,\ref{FigureCDW}{\bf d-f}. Tunneling conductance maps at different bias voltages are provided in Extended Data Fig.\,\ref{FigureBias} and in Supplementary videos. We identify Bragg peaks associated with the atomic lattice, marked with white circles, and an additional set corresponding to the CDW. The surface Brillouin zone is delineated by a red hexagon in Fig.\,\ref{FigureCDW}{\bf d-f}. The CDW peaks indicated by orange circles are consistent with those reported in previous studies\cite{Aishwarya2023,Gu2023,Aishwarya2024,LaFleur2024}. Notably, two peaks along the long axis of the hexagon are located clearly outside the first Brillouin zone of the surface. The remaining four peaks reside close but outside the Brillouin zone boundary. In addition, we observe Bragg peaks, indicated by yellow circles, which are located at approximately half the wavevectors of previously reported Bragg peaks. We attribute these newly observed Bragg peaks to the primitive CDW wavevectors, determined to be (along the surface unit cell axes defined in Methods) $\boldsymbol{Q}_{CDW1}=(0.72\pm0.08,0.02\pm0.08)$ nm$^{-1}$ and $\boldsymbol{Q}_{CDW2}=(0.53\pm0.08, 0.34\pm0.08)$  nm$^{-1}$.

The electronic density of states spatially modulated by the CDW can be expressed as $N_L(\boldsymbol{r})\propto N_0 \sin(\boldsymbol{Q}_{CDW}\cdot\boldsymbol{r}+\varphi_{CDW})$, with $N_0$ the amplitude of the modulations and $\varphi_{CDW}$ the phase. We find that the phase $\varphi_{CDW}$ is independent in different terraces (Supplementary Information Section 2 and Extended Data Figs.\,\ref{FigurePhase1},\ref{FigurePhase2},\ref{FigurePhase3}). This suggests that the CDW has a strongly two-dimensional character and is localized at the surface plane.

\paragraph*{Band structure of UTe$_2$.} 
To elucidate the role of the electronic band structure in the CDW formation, we examine the Fermi surface derived from density functional theory (DFT) incorporating Coulomb repulsion $U$ (Fig.\,\ref{FigFS}{\bf a}, for $U=$ 2 eV). The Fermi surface is predominantly composed of two nearly two-dimensional tubular bands situated within the $k_x$-$k_y$ plane, exhibiting a slight warping along the $k_z$-axis. These bands display electron-like character along the $k_x$ axis and hole-like character along the $k_y$ axis\cite{Ishizuka2019,shimizu2025}. The U 5f orbital weight dominates over the whole Fermi surface, with little directional dependence (Extended Data Fig.\,\ref{FigBandstructure}{\bf b}). These calculations reproduce well the Fermi surface geometry, although they fail to account for the exceptionally large effective masses observed in UTe$_2$ via de Haas van Alphen and specific heat measurements\cite{
Aoki_dHvA2022,Aoki_dHvA2023,Eaton2024,AokiReview2022}, consistent with results in other heavy fermion metals\cite{Taillefer88}.

Despite the limitations in reproducing effective masses, it is useful to analyze the electronic properties expected for the (011) surface. These surface properties are determined by the projection of all isoenergy planes parallel to the surface, i.e.\,planes perpendicular to ${\bm k}_{\perp}$, the vector normal to the surface plane, ranging from ${\bm k}_{\perp}=0$ to ${\bm k}_{\perp}=\Gamma-S$, where $\Gamma-S$ is the vector joining the center $\Gamma$ and the $S$ point of the bulk Brillouin zone (Fig.\,\ref{FigureIntro}{\bf d,e}). This is due to the fact that ${\bf k}_{\perp}$ is not a good quantum number at the surface. The intersections of the bulk Brillouin zone with planes at $k_{\perp}=0$ and $k_{\perp}=|\Gamma-S|$, i.e. through the $\Gamma$ and $S$ points, are shown by grey rectangles in Fig.\,\ref{FigFS}{\bf a}. The corresponding Fermi surface contours for the $k_{\perp}=|\Gamma-S|$ plane are shown in Fig.\,\ref{FigFS}{\bf b}. To identify the $k_{\perp}$ values which contribute most significantly to the tunneling conductance, we calculated the normalized two-dimensional density of states in each plane, using the formula $g_{Norm,\nu}(k_{\perp})=\frac{1}{L}\int_L\frac{dL}{\vert \nabla_{\bm k} E({\bm{k}})\vert }$, where $L$ is the Fermi surface contour length, $E(\bm{k})$ the full three-dimensional energy dispersion and $\nu$ the band index (see Extended Data Fig.\,\ref{FigContourIntegral} and Supplementary Information, Section 3 for more details). The results for $g_{Norm,\nu}(k_{\perp})$ for the electron and hole-like bands are shown in Fig.\,\ref{FigFS}{\bf c}. The largest contributions originate from the planes at the Brillouin zone center ($k_{\perp}=0$) for the hole-like band and boundary ($k_{\perp}=|\Gamma-S|$) for the electron-like band. At $k_{\perp}=|\Gamma-S|$, the Fermi surface contour comprises two parallel lines from the electron-like band. The arrow in Fig.\,\ref{FigFS}{\bf b} represents the $\boldsymbol{Q}_{CDW1}$ wavevector observed in our experiment. The near coincidence of this wavevector with the parallel lines suggests that the CDW wavevector along this direction connects near-parallel features in the electron-like band. To discuss features associated with the $\boldsymbol{Q}_{CDW2}$ wavevector, we must consider the Fermi contour at $k_{\perp}=0$, which primarily intersects the hole-like band.

The Fermi surface contour of the hole-like band within the $k_{\perp}=0$ plane is shown in Fig.\,\ref{FigFS}{\bf d} and exhibits significantly greater complexity than the two parallel lines depicted in Fig.\,\ref{FigFS}{\bf b}, extending well beyond the boundaries of the surface Brillouin zone (indicated in red in Fig.\,\ref{FigFS}{\bf b,d}). To represent the two-dimensional Fermi surface contour, we performed a folding operation, mapping the $k_{\perp}=0$ hole-like band contour into the first surface Brillouin zone by vector addition of reciprocal lattice wavevectors of the surface lattice, denoted as $\overline{\bm{G}}_{S}$, to contour segments located outside the first Brillouin zone. As shown in Fig.\,\ref{FigFS}{\bf d} we find two vectors connecting band structure segments, oriented at an angle relative to the $k_x$-axis. We emphasize that these vectors arise only at the surface after downfolding the band structure. These vectors (indicated by dark green arrows in Fig.\,\ref{FigFS}{\bf d}) are equivalent by symmetry and coincide with the wavevector $\boldsymbol{Q}_{CDW2}$ observed in our experiments.

These Fermi surface features are distinct from those associated with enhanced antiferromagnetic fluctuations observed in neutron scattering experiments. The latter occur at a wavevector (0,$\pm$0.93,0) nm$^{-1}$\cite{PhysRevB.104.L100409,PhysRevLett.125.237003,Butch2022}. But this wavevector has no relation to $\boldsymbol{Q}_{CDW1}$ or $\boldsymbol{Q}_{CDW2}$, even if we consider just its projection to the surface plane (which gives $(\overline{k_x}$, $\overline{k_y})=(0,0.37)\ \textrm{nm}^{-1}$). Thus, the feature at $\boldsymbol{Q}_{CDW1}$ does not have a significant influence in the known bulk properties, and the one at $\boldsymbol{Q}_{CDW2}$ is only found at the surface as it arises from a feature obtained by downfolding the surface Brillouin zone. Both $\boldsymbol{Q}_{CDW1}$ and $\boldsymbol{Q}_{CDW2}$ are discernible features of the Fermi surface in DFT calculations, and present strong nesting, thus favoring interactions at these wavevectors.

\paragraph*{CDW and the heavy fermion density of states in UTe$_2$.} 

A phonon-mediated CDW appears in $\alpha-$U (see Supplementary Information, Section 4). However, in rare earth compounds, CDW formation is found to compete with Kondo screening\cite{Torikachvili07,Trontl,Derr06}. CDWs are indeed relatively unfrequent in heavy fermion systems. This, together with the absence of indications for a CDW from structural studies in UTe$_2$\cite{kengle2024absence,kengle2024absence2} and the distinct heavy masses not reproduced by DFT\cite{
Aoki_dHvA2023,Eaton2024,AokiReview2022}, prompts us to question the eventual modifications suffered by the electronic band structure at the surface.

The heavy fermion quasiparticles lead to a dip-peak structure in the tunneling conductance of UTe$_2$, which is due to co-tunneling into heavy and light quasiparticles and the hybridization gap opening\cite{Jiao2020,Morr17,Hamidian11}. In Fig.\,\ref{Steps}{\bf a} we show the tunneling conductance on flat terraces in UTe$_2$, together with a fit to the usual expression for heavy fermion low energy hybridization, described in the Methods section. We find that the shape of the dip-peak structure in the tunneling conductance changes together with the CDW (light and dark green crosses on the inset of Fig.\,\ref{Steps}{\bf a}), being compatible with local changes in the hybridization strength by about 10\% (Extended Data Table\,\ref{TableTerrace} and Methods).
 
To explore further the role of the surface in the heavy fermion hybridization, let us consider the U\,5f-electron character of the band structure at large energies (Extended Data Fig.\,\ref{FigBandstructure}). A few hundred of meV below the Fermi level, the density of states is dominated by a peak from U\,6$\mathrm{d_{z^2}}$ orbitals, reminiscent of the one-dimensional character of this band and compatible with a strong increase of the tunneling conductance for tunneling into filled states observed in our data (Extended Data Fig.\,\ref{FigBandstructure}{\bf a}). Generally, surface steps lead to a shift in the band structure due to uncompensated charges at the step edges (see Supplementary Information, Section 5 and Extended Data Fig.\,\ref{FigSpaceFillingPolyhedra} for more details). We observe that at surface steps the tunneling conductance is considerably modified and a strong peak, centered around -200\,meV, arises (Extended Data Fig.\,\ref{FigBandstructure}{\bf a}). This suggests that the bottom of the U\,6$\mathrm{d_{z^2}}$ derived band (Extended Data Fig.\,\ref{FigBandstructure}{\bf d}) shifts towards the Fermi level at the step edge. The shift in the U\,6$\mathrm{d_{z^2}}$ band structure drives a decrease in the f-electron valence towards U$^{4+}$\cite{PhysRevB.19.6615,Christovam24}. This should modify considerably the heavy fermion hybridization. When we now look at the tunneling conductance at small bias voltages on the step edges, we find distinct curves as compared to flat terraces. We observe a decrease of the size of the characteristic dip-peak structure of UTe$_2$ (Fig.\,\ref{Steps}), with a heavy fermion hybridization gap which decreases down to well below 1\,meV (Table\,\ref{TableTerrace}). The valence state of rare earth and actinide elements is indeed known to vary at surfaces, typically shifting towards a lower f-electron count\cite{PhysRevB.19.6615}. The shift increases on steps and even more on corners. Experimental determinations of the U 5f valence in UTe$_2$ using photoemission indicate a tendency towards 5f occupancy near two (U$^{4+}$) for surface-sensitive and near three  (U$^{3+}$) for bulk-sensitive measurements\cite{PhysRevLett.124.076401,FujimoriJPSJ2019,Liu22,Christovam24}. The behavior at steps observed here directly shows the change in the f-electron valence when exposing UTe$_2$ to vacuum. This tendency, although less pronounced, should affect atomically flat surfaces too. We find a Kondo temperature of about 20\,K (Fig.\,\ref{Steps} and Extended Data Table\,\ref{TableTerrace}) for flat surfaces, compatible with previous surface sensitive measurements\cite{Jiao2020,PhysRevLett.124.076401,FujimoriJPSJ2019,Liu22,Christovam24}. Furthermore, as we show in Fig.\,\ref{Steps}, the hybridization strength changes characterize the spatial modulation of the electronic density at low energies induced by the CDW, suggesting a new form of heavy fermion charge order.

The concept of a spatial variation of the hybridization leading to periodic charge oscillations has been theoretically explored and experimentally observed, with charge modulations at nesting wavevectors that are close to those of the high energy unhybridized band structure. However, these remained local around Kondo holes and impurities\cite{Morr17,Hamidian11}. The occurrence of a spatially homogeneous CDW at the surface of UTe$_2$ could be related to the slight discrepancy obtained between $\boldsymbol{Q}_{CDW1}$ and the calculated nesting wavevector mentioned above. Dynamical many-body effects could renormalize the low-energy band structure and redistribute the orbital character on the Fermi surface, thereby modifying the detailed momentum dependence of the electronic susceptibility. Spatially extended hybridization waves were theoretically proposed within the periodic Anderson model and as a novel form of low temperature electronic ordering in the hidden order state of URu$_2$Si$_2$\cite{Kawakami2011,Dubi2011,Hafner22,Huang19}.

\paragraph*{Conclusion.} Leveraging enhanced resolution at low wavevectors in high-resolution STM experiments, we have determined the primitive wavevectors of the CDW in UTe$_2$ and demonstrated that they are located well within the surface Brillouin zone. We also established that CDW occurs only at the surface. By comparing our experimental results with band structure calculations that accurately reproduce the Fermi surface geometry, we have identified two wavevectors providing enhanced interactions and matching the primitive wavevectors of the CDW. Furthermore, the behavior observed at steps suggests that surface-induced modifications in the valence of Uranium f-electrons drives a reduction of the Kondo temperature and allows the establishment of a periodic modulation of the hybridization pattern.

The Fermi surface features at  $\boldsymbol{Q}_{CDW1}$ and  $\boldsymbol{Q}_{CDW2}$ do not produce charge or spin ordering in the bulk. However, they are visible in several experiments addressing both surface and bulk properties. For example, very recent experiments have found modulations of the in-gap superconducting density of states at some of the primitive CDW wavevectors~\cite{Wang25}, adding to the previously observed superconducting PDW associated to the CDW~\cite{Gu2023}. Furthermore, calculations hint at possible nodes in the superconducting order parameter at the primitive CDW wavevectors~\cite{pttl-8m71,3kcp-w4ny}. Additionally, the superconducting coherence length anisotropy leads to strongly elongated vortex cores, compatible with the nearly one-dimensional character of the band giving $\boldsymbol{Q}_{CDW1}$~\cite{shimizu2025,Yin25,Sharma25,Yang25}. This suggests that the Fermi surface hot spots found here might also have a profound influence in the bulk properties of UTe$_2$.

\clearpage

\begin{figure}
\begin{center}
	\includegraphics[width = \textwidth]{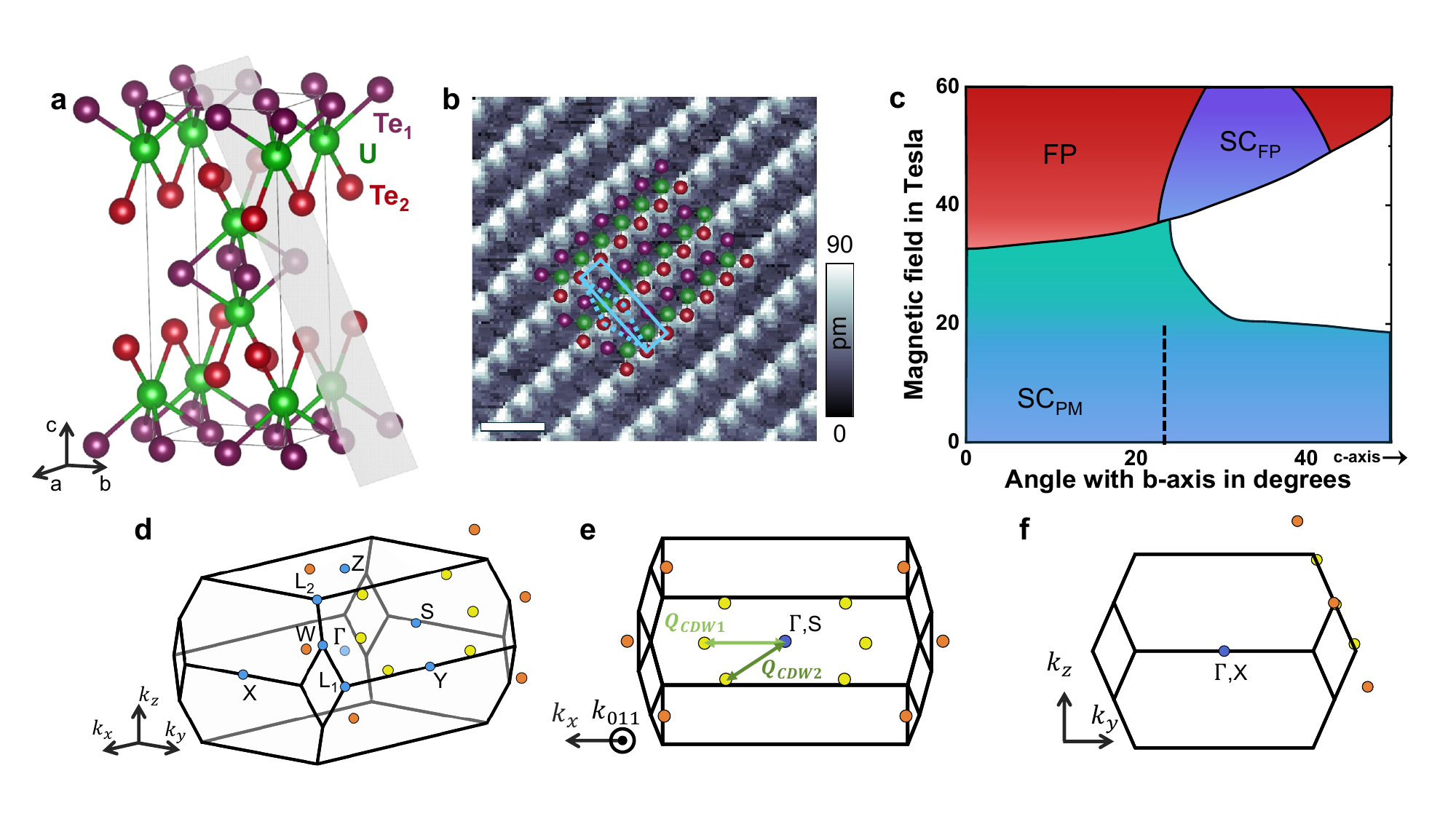}
\end{center}
	\caption{\noindent {\bf \,| Crystal structure and CDW of UTe$_2$.} {\bf a} Atomic lattice of UTe$_2$. Violet, red and green spheres depict Te(1), Te(2) and U atoms, respectively. The cleaved (011) plane is given in grey. {\bf b} Atomic resolution topographic image of the UTe$_2$ surface. The atomic lattice positions are marked by dots, with colors corresponding to those in {\bf a}. The white scale bar represents 1 nm. Te(2) atoms form chains along the crystallographic a-direction. The rectangle shown with solid blue lines defines the surface unit cell, while the primitive crystal unit cell is indicated by dashed blue lines. {\bf c} Schematic phase diagram of UTe$_2$, adapted from Refs.\,\cite{Ran2019,Lewin_2023}. Phases are labelled as field polarized (FP), superconducting in the FP phase (SC$_{FP}$) and superconducting in the paramagnetic phase (SC$_{PM}$). The black dashed line indicates the direction of the applied magnetic field, perpendicular to the cleaved (011) surface. {\bf d-f} Black lines schematically represent the bulk Brillouin zone. Reciprocal space axes are shown in the bottom left corner of each panel. Blue dots denote selected high symmetry points of the bulk Brillouin zone. The complete set of high-symmetry points for the bulk and the surface Brillouin zones is presented in Extended Data Fig.\,\ref{FigBZs} and discussed in Supplementary Information, Section 1. Orange dots provide the positions of the Bragg peaks previously reported in Refs.\,\cite{Aishwarya2023,Gu2023,Aishwarya2024,LaFleur2024}. Yellow dots denote the position of the Bragg peaks of the CDW identified in this study, and the green arrows represent the corresponding wavevectors $Q_{CDW1}$ and  $Q_{CDW2}$.}
    \label{FigureIntro}
\end{figure}

\begin{figure}[t]
		\centering
		\begin{center}
			\includegraphics[width = 0.9\textwidth]{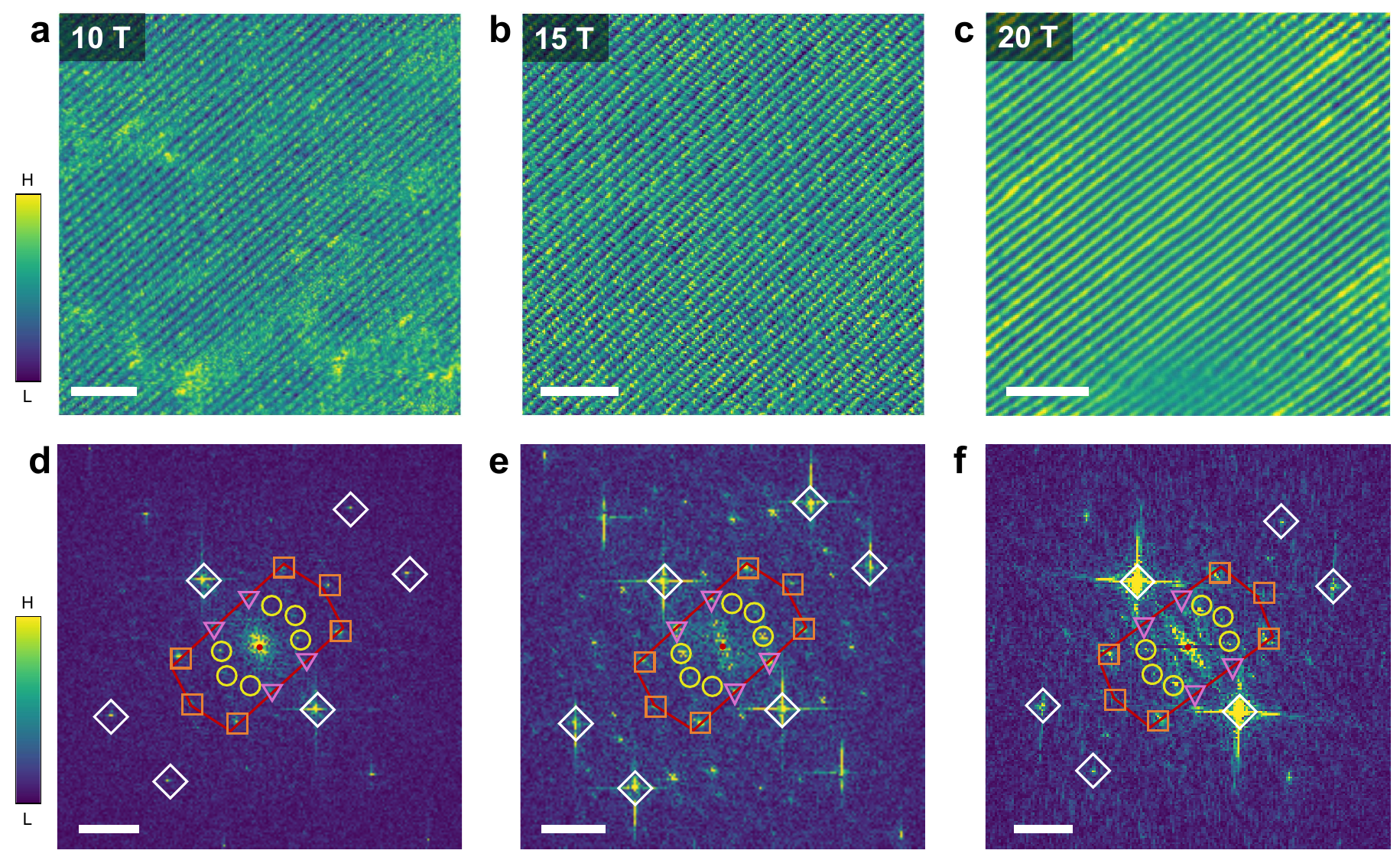}
		\end{center}
	\caption{\noindent{\bf \,| Tunneling conductance maps of the CDW.} Tunneling conductance maps of UTe$_2$ at 10 T {\bf a}, 15 T {\bf b}, and 20~T {\bf c}, acquired from three distinct fields of view, and their corresponding Fourier transforms {\bf d,e,f}. The white scale bars are 5 nm long. The bias voltage for all three maps is close to zero. The amplitude for the different peaks varies with bias voltage. For tunneling conductance maps as a function of the bias voltage, refer to Extended Data Fig.\,\ref{FigureBias} and Supplementary videos. Bragg peaks associated with the atomic lattice are marked by white rhombuses in {\bf d,e,f}. The surface Brillouin zone is delineated by red lines in {\bf d,e,f} (see also Extended Data Fig.\,\ref{FigBZs} and Supplementary Information, Section 1). Bragg peaks corresponding to charge modulations reported in previous experiments are located slightly outside the surface Brillouin zone boundary and are marked by orange squares. The newly identified CDW wavevectors are indicated by yellow circles in {\bf d,e,f}. Not all peaks are observed on the maps shown here close to zero bias. The other positions arise at other bias voltages. All other peaks observed in the maps, as for example the triangles marked in violet, are linear combinations of the primitive CDW wavevectors identified here (see Supplementary Information, Section 6). The white scale bar in the Fourier transforms {\bf d-f} represents 1 nm$^{-1}$. The color scale for each map is indicated by the bars on the left.}
  \label{FigureCDW}
\end{figure}

\begin{figure}[t]
		\centering
		\begin{center}
			\includegraphics[width = 0.80\columnwidth]{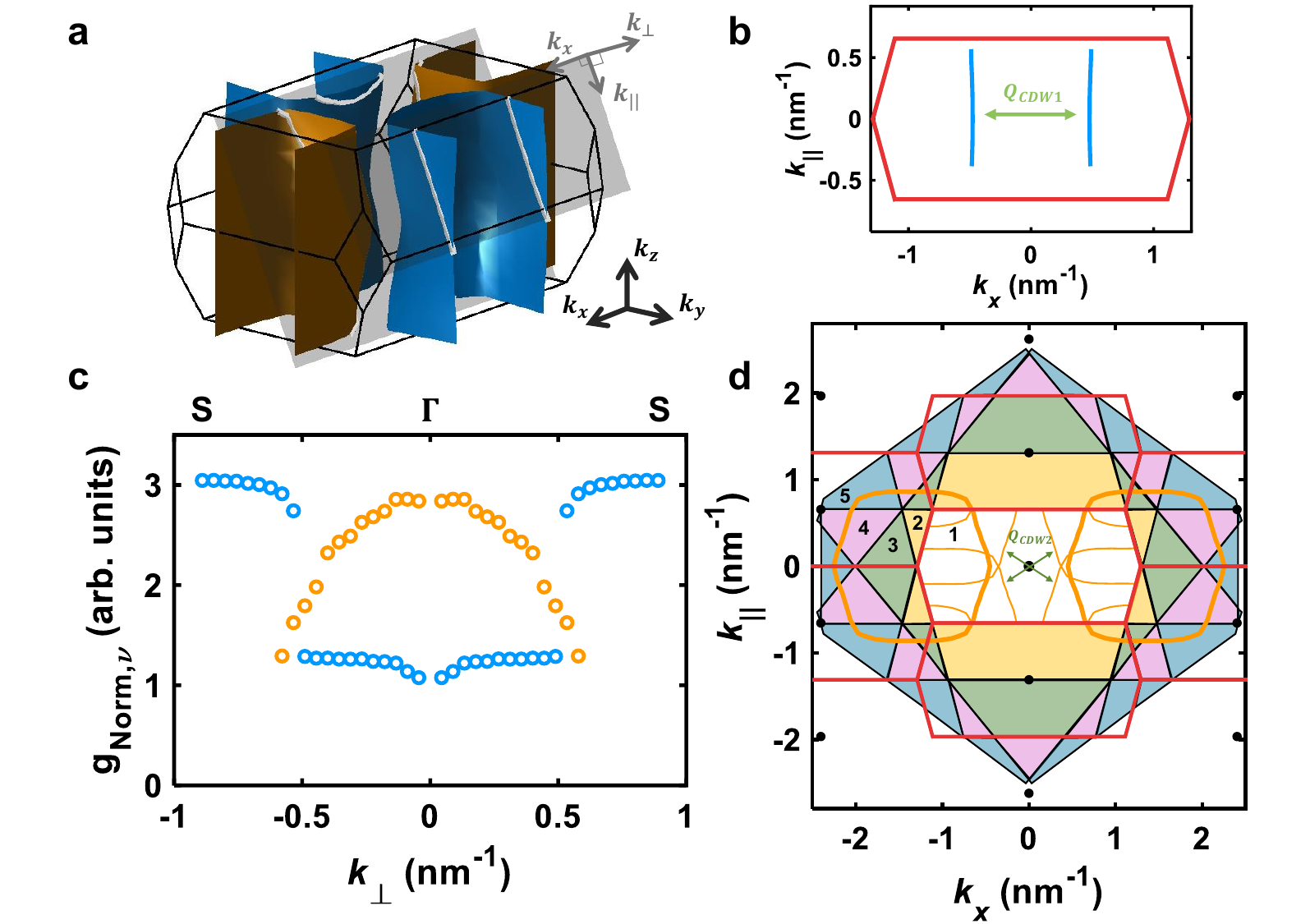}
		\end{center}
		\caption{\noindent{\bf \,| Bulk Fermi surface and Fermi contours at the surface of UTe$_2$.} {\bf a} The Fermi surface is represented as a colored contour, with blue indicating electron-like bands and orange indicating hole-like bands. The bulk Brillouin zone is delineated by black lines. Reciprocal space planes perpendicular to ${\bm k}_{\perp}$, specifically at $k_{\perp}=0$ and $k_{\perp}=|\Gamma-S|$, are shown in grey. As shown in the Extended Data Fig.\ref{FigBandstructure}{\bf b}, the whole Fermi surface has a pronounced U 5f character, which dominates in the features shown in {\bf b,c,d}. {\bf b} Fermi surface contour at the reciprocal space plane $k_{\perp}=|\Gamma-S|$. The  ${\bm k}_{\parallel}$ axis is perpendicular to ${\bm k}_x$ within the (011) surface plane in reciprocal space (depicted in grey in {\bf a}). The surface Brillouin zone is delineated by a red line (see also Extended Data Fig.\,\ref{FigBZs}). The wavevector $\bm{Q}_{CDW1}$, indicated by a light green arrow, is very close to the nesting wavevector of the electron-like band. {\bf c}  Normalized density of states of the Fermi surface contour $g_{Norm,\nu}({\bm k}_{\perp})=\frac{1}{L}\int_L\frac{dL}{\vert \nabla_{\bm{k}} E\vert }$, where $L$ is the length of the Fermi surface contour in reciprocal space and $\nu$ is the band index. Blue and orange colors correspond to the respective band, as in {\bf a}. Top axis indicates the high symmetry points that the plane passes through. {\bf d} We depict the atomic Bragg lattice as black points within the reciprocal space surface plane ${\bf k}_{\perp}=0$. The first Brillouin zone of the surface is delineated by red lines and is repeated into adjacent reciprocal space regions. The Fermi contour of the hole band is represented by orange lines. The surface Brillouin zone order is numbered from $1-5$ and the zones are distinguished by different colors. The Fermi contours in higher-order surface Brillouin zones are folded into the first surface Brillouin zone by vector addition of a reciprocal lattice wavevector of the surface $\overline{{\bm G_S}}$. Dark green arrows indicate the experimentally determined wavevector $\bm{Q}_{CDW2}$.}
  \label{FigFS}
\end{figure}

 \begin{figure}[t]
		\centering
		\begin{center}
			\includegraphics[width = 0.98\columnwidth]{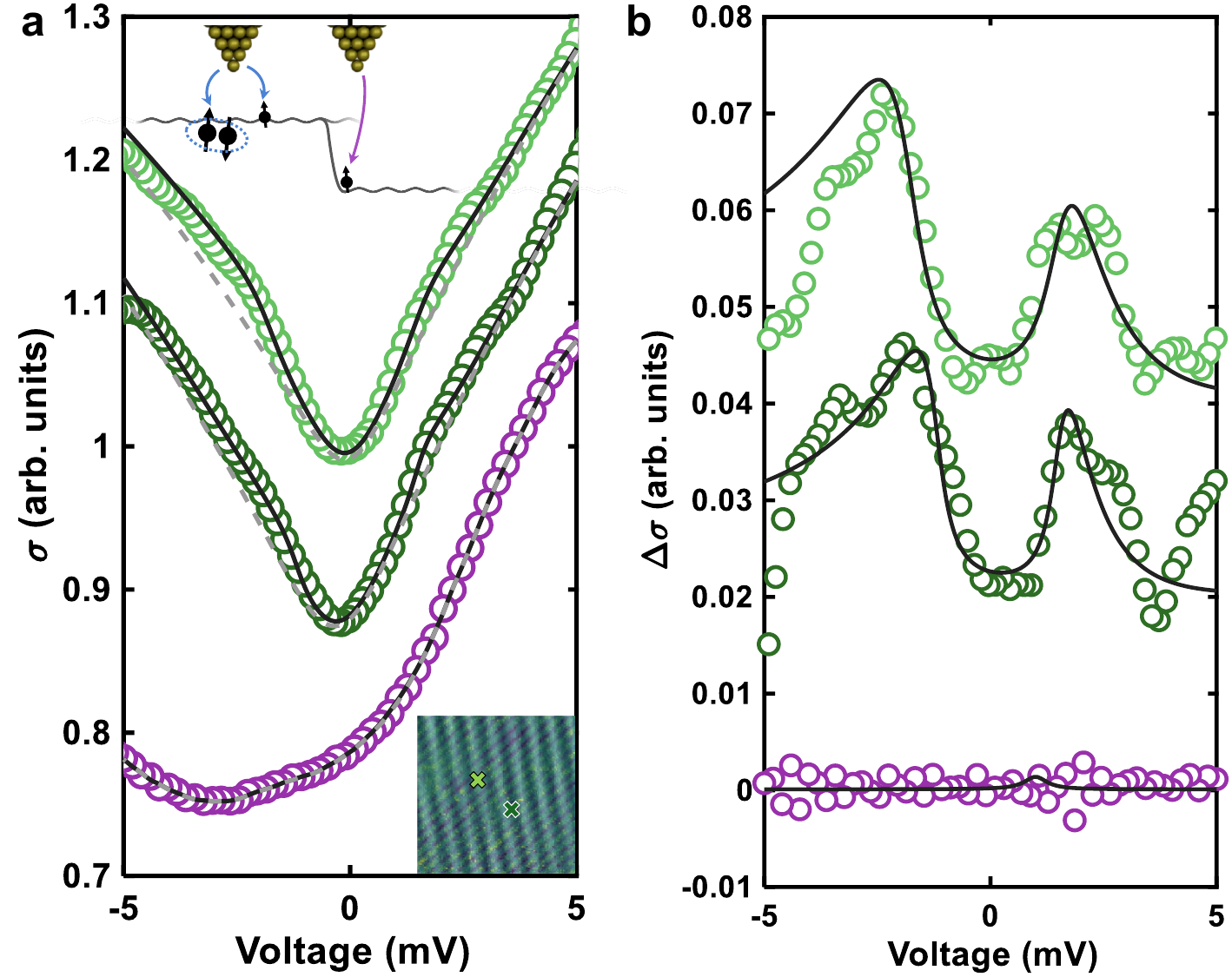}
		\end{center}
		\caption{\noindent{\bf \,| Hybridization wave in UTe$_2$ and reduction of heavy fermion hybridization close to a step.} {\bf a} Light and dark green circles show the tunneling conductance measured on top of an atomically flat area at two different positions marked by crosses with the same colors in the bottom right inset. The fits to the heavy fermion co-tunneling model described in the Methods section are shown as black lines (parameters given in Extended Data Table\,\ref{TableTerrace}). In the bottom right inset we show a tunneling conductance map within an atomically flat terrace, acquired at -2mV, which highlights the Fourier filtered tunneling conductance modulations at $\bm{Q}_{CDW2}$. Te(2) chains show as lines going from bottom left to top right. We also show as violet colored points the tunneling conductance obtained on a step edge. The fit to the cotunneling model is shown by a black line (parameters given in Extended Data Table\,\ref{TableTerrace}). Top inset provides a schematic representation of the tunneling processes depicted in the main panel. As shown by the violet line, cotunneling is strongly weakened at the step due to the change in f-electron valence (see also Extended Data Fig.\,\ref{FigBandstructure} and Supplementary Information, Section 5). {\bf b} We show the same data as colored circles, with a V-shape background subtracted (the background is shown by a grey dashed line in {\bf a}).}
  \label{Steps}
\end{figure}

\clearpage

\section*{Methods}

\paragraph*{Scanning Tunneling Microscopy. }
We employed a custom-built millikelvin STM, incorporating in-house developed STM electronics and software, as described in Refs.\cite{marta2021,fran2021}. The STM was mounted into a dilution refrigerator and the bore of a 20\,T magnet, both manufactured by Oxford Instruments, except the gas handling system which was in-house developed. Atomic resolution imaging and tunneling spectroscopy, achieving an energy resolution below 8\,$\upmu$eV, have been demonstrated in different materials using this apparatus\cite{marta2021}. Data processing was performed using custom software developed in-house\cite{Githublbtuam}, supplemented by standard image visualization software\cite{Horcas07}. The data presented herein were acquired at the base temperature of approximately 80\,mK, a value previously validated by measuring the superconducting properties of aluminium\cite{marta2021}. The single crystals of UTe$_2$ were obtained using the chemical vapor transport method\cite{AokiJPSJ2020}. Sample cleavage was performed following the procedures outlined in Refs.\cite{Herrera2021,Herrera2023}. UTe$_2$ crystals were oriented and pre-cut to produce rectangular specimens with a long axis perpendicular to the (011) crystallographic plane and a minimized cross-sectional area. The sample and a gold pad were glued to a copper holder using silver epoxy. A string was attached to the upper section of the sample. Below liquid helium temperatures, the sample holder was manipulated using the system described in Ref.\,\cite{suderow2011} to direct the sample towards a beam. This resulted in sample fracturing into two parts. One part was guided to the bottom of the vacuum chamber using the string, while the other retained a freshly cleaved (011) surface. The STM tip was initially approached to the gold surface for conditioning, employing the method described in Ref.\,\cite{Rodrigo04b}. Following verification of atomic sharpness, the sample holder was repositioned to allow for the study of the cleaved (011) surface of UTe$_2$. Multiple fields of view, each approximately $2\times2$ $\upmu$m in size, were investigated. The results presented in this study represent a compilation of data acquired from over one hundred distinct fields of view.

To determine the three dimensional coordinates of the CDW wavevectors, we utilize the unit vector normal to the (011) surface, $\hat{\bm{k}}_{\perp} = \frac{\bm{k}_{\perp}}{k_{\perp}} = \frac{(0,c,b)}{\sqrt{b^2+c^2}}$ and  two orthogonal unit vectors within the surface plane, $\hat{\bm{k}}_x = (1,0,0)$ and $\hat{\bm{k}}_{||} = \hat{\bm{k}}_{\perp} \times \hat{\bm{k}}_x$. The vectors $\hat{\bm{k}}_x$ and $\hat{\bm{k}}_{||}$ define the in-plane coordinate system of the (011) surface in reciprocal space. We calculate the in-plane components of the positions of the CDW within the three dimensional bulk Brillouin zone by projecting the vectors providing the positions of the CDW peaks on the surface plane, using ($\overline{k_x}$, $\overline{k_y}$) =$(\hat{\bm{k}}_x \cdot \bm{Q}_i, \hat{\bm{k}}_{||} \cdot \  \bm{Q}_i)$, where $\bm{Q}_i$ represent the positions of the CDW peaks in the bulk three-dimensional Brillouin zone\cite{kengle2024absence}. We estimate the error in the wavevector from the width of the CDW peaks in reciprocal space.

The additional CDW Bragg peaks we observe here are also discernible in previous work, for example in Ref.\,\cite{Aishwarya2023} but were likely overlooked. This oversight may have resulted from the obscuring influence of density of states fluctuations at small wavevectors, potentially induced by defects leading to a decreased resolution for small wavevectors. As detailed in the main text, the Bragg peaks observed in this study are located entirely within the first Brillouin zone of the bulk and surface lattices. In contrast, previously observed Bragg peaks are situated well outside the bulk Brillouin zone and marginally outside the surface Brillouin zone.

\paragraph*{Band structure calculations. }We performed fully relativistic all-electron density functional theory (DFT) calculations for high-quality UTe$_2$~\cite{Haga2022, Sakai2022} using the full potential local orbital (FPLO) basis~\cite{Koepernik1999} and the generalized gradient approximation (GGA) exchange-correlation functional~\cite{Perdew1996}. To account for the electronic correlations we employed GGA + $U$ calculations with the around mean-field double-counting correction scheme~\cite{Ylvisaker2009}. A $12 \times 12 \times 12$ $k$-point mesh was utilized. In this study, we select Slater parameters for U $5f$ orbitals as $(F_0, F_2, F_4, F_6) = (2\mathrm{\;eV}, 0, 0, 0)$, corresponding to $\tilde{U} = U - J = 2\mathrm{\;eV}$. With these parameters, we obtain a cylindrical Fermi surface characterized by significant U-$5f$ orbital contribution at the Fermi level, consistent with previous DFT studies and with the geometry of the Fermi surface obtained from de Haas van Alphen and angle resolved photoemission measurements~\cite{Ishizuka2019,Harima2020,Aoki_dHvA2022,Aoki_dHvA2023,Broyles2023,Eaton2024,Weinberger2024,Miao2020}.   To extract the electronic properties on the surface plane we imported the FPLO output files into Matlab and defined the surface plane and associated cuts using code available in Ref.\,\cite{Githublbtuam}.

\paragraph*{Co-tunneling in UTe$_2$.} The tunneling conductance between a tip of a normal metal and a heavy fermion has been discussed with detail in literature\,\cite{Jiao2020,Morr17,Hamidian11}. Tunneling into a single Kondo impurity in an otherwise normal metal occurs through two interfering channels, consisting of tunneling into the virtual Kondo bound state and into the conduction electrons, and leads to an asymmetric Fano shaped tunneling conductance\cite{Morr17,Wolfle10,Figgins10}. In a heavy fermion, the low temperature hybridization between light carriers and localized levels leads to an avoided crossing close to the Fermi level and the formation of heavy quasiparticles. In addition to the Fano feature, the tunneling conductance presents gap opening. The tunneling conductance  as a function of the bias voltage is then described by an expression of the form\cite{Maltseva09,Jiao2020,Morr17,Wolfle10,Figgins10}:

\begin{equation*}
\sigma(V) \propto \operatorname{Im}\left[ \left( 1+\frac{\nu_K q}{eV-i\Gamma-E_r} \right)^2 \log \left( \frac{eV-i\Gamma+E_1-\frac{\nu_K^2}{eV-i\Gamma-E_r}}{eV-i\Gamma-E_2-\frac{\nu_K^2}{eV-i\Gamma-E_r}} \right) + \frac{D q^2}{eV-i\Gamma-E_r} \right],
\end{equation*}

where $D$ is the unrenormalized bandwidth (with -$E_1$ and $E_2$ being the bottom and top of the band), $q$ the ratio between tunneling amplitudes into the heavy and the light electron states, $\Gamma$ a broadening parameter, $E_r$ the resonance energy, and $\nu_K$ the hybridization strength. To fit the tunneling conductance observed in our experiment (Fig.\,\ref{Steps}) we adapt the parameter set used in previous work (Refs.\,\cite{Jiao2020,Hamidian11}) to account for our band structure calculations. We use $D, E_1$ and $E_2$ compatible with the band shown in Extended Data Fig.\,\ref{FigBandstructure}{\bf d} and find the parameters shown in Extended Data Table\,\ref{TableTerrace}. We note that in the two-unit cell step analyzed in Fig.\,\ref{Steps}, Extended Data Fig.\,\ref{FigBandstructure}{\bf a} and Extended Data Fig.\,\ref{FigSpaceFillingPolyhedra}, the hybridization is considerably modified.

\paragraph*{Acknowledgments.} \
\noindent We acknowledge discussions with J.P. Sanchez and with G. Lander. Support by the Spanish Research State Agency (PID2023-150148OB-I00, TED2021-130546B-I00, PDC2021-121086-I00 and CEX2023-001316-M), the European Research Council PNICTEYES through grant agreement 679080 and VectorFieldImaging Grant Agreement 101069239, the EU through grant agreement No. 871106, by the Comunidad de Madrid through projects TEC-2024/TEC-380 “Mag4TIC” and PhD thesis support PIPF-2023/TEC-30683 and PIPF-2023/TEC-30853 are acknowledged. M.A.V., E. H., I.G and H. S. acknowledge the QUASURF project (ref. SI4/PJI/2024-00199) funded by the Comunidad de Madrid through the agreement to promote and encourage research and technology transfer at the Universidad Aut\'onoma de Madrid. We have benefited from collaborations through EU program Cost CA21144 (superqumap), and from SEGAINVEX at UAM in the design and construction of STM and cryogenic equipment.

\paragraph*{Author contributions.} \
\noindent  P.G.T. performed the experiments with the supervision of I.G. and E.H., and support by M.A.V., P.G.T. and M.A.V. worked together and with M.S. and O.B.M. on data treatment and analysis. M.S. and Y.Y. performed the band structure calculations. M.A.P., G.K., D.A., M.A.V. and G.L. provided samples and helped to mount these in the STM, together with B.W., G.K., J.F. and J.P.B., Y.Y. provided guidance about the properties of UTe$_2$ and the interpretation of the results and co-supervised the work, with E.H., I.G. and H.S. who wrote the manuscript, together with P.G.T. All authors contributed to the manuscript.

\paragraph*{Data availability.} All data and calculations are available upon request.
\noindent 

\noindent 

\clearpage

\setcounter{table}{0}

\captionsetup[table]{labelfont={bf},name={Extended Data Table \let\nobreakspace\relax},labelsep=none}

\begin{table}
\centering
\begin{tabular}{|c|c|c|c|c|c|c|c|}
\hline
\textbf{Location} & $\bm{q}$ & $\bm{E_r}$ \textbf{(meV)} & $\mathbf{\Gamma}$ & $\bm{\nu_K}$ \textbf{(meV)} & $\bm{E_1}$ \textbf{(eV)} & $\bm{E_2}$ \textbf{(eV)} & $\bm{T_K}$ \textbf{(K)} \\
\hline
CDW peak & -0.25 & 0.4 & 0.5 & 21 & 0.4 & 0.2  & 17.1\\
\hline
CDW trough & -0.25 & 0.6 & 0.3 & 19 & 0.4 & 0.2 & 14.9 \\
\hline
Step & -0.14 & 1 & 0.3 & 0.2 & 0.4 & 0.2  & 0.001\\
\hline
 \end{tabular}
	\caption{\noindent {\bf \,|  Co-tunneling density of states on the CDW and on a step.} Parameters obtained by fitting the tunneling conductance to the expression shown in Methods. We have adjusted the bottom and top of the unrenormalized band, -$E_1$ and $E_2$, to the values provided by the DFT calculations, see Extended Data Fig.\,\ref{FigBandstructure}{\bf d}. $E_r$ is the energy position of the f-electron resonant level, $q$ the ratio of tunneling amplitudes on heavy and light electrons and $\nu_K$ the Kondo hybridization strength. The hybridization strength is related to the Kondo temperature through $T_K\approx \frac{1}{k_B}\frac{2\nu_K^2}{(E_1+E_2)}$\cite{Maltseva09,Jiao2020,Morr17,Wolfle10,Figgins10}. We observe that $\nu_K$ strongly decreases on a step edge. Within an atomically flat terrace, however, $\nu_K$ fluctuates together with the CDW.}
 \label{TableTerrace}
\end{table}

\clearpage

\setcounter{figure}{0}

\captionsetup[figure]{labelfont={bf},name={Extended Data Figure \let\nobreakspace\relax},labelsep=none}

\begin{figure}
\begin{center}
	\includegraphics[width=0.98\textwidth]{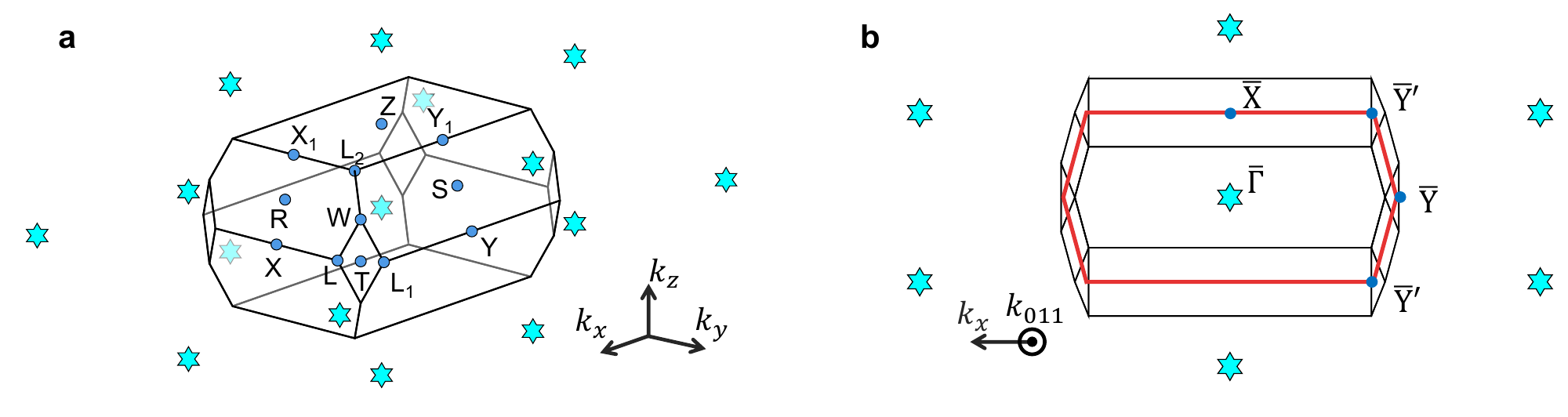}
\end{center}
	\caption{\noindent {\bf \,|  Surface (011) and bulk Brillouin zones of UTe$_2$.} {\bf a} The bulk Brillouin zone of UTe$_2$ is delineated by black lines. The zone center, designated as the $\Gamma$ point, is indicated by a blue star. Additional high-symmetry points are also marked. The positions of the $\Gamma$ points in all 14 adjacent bulk Brillouin zones are also represented by blue stars. {\bf b} The bulk Brillouin zone, as in {\bf a}, projected into the (011) surface plane is shown by black lines. The surface Brillouin zone, delineated by red lines, corresponds to the Wigner-Seitz cell of the surface reciprocal lattice, defined by projected bulk $\Gamma$ (blue stars). High symmetry points of the surface Brillouin zone are indicated by blue points and labeled as $\overline{\Gamma}$, $\overline{X}$,  $\overline{Y}$ and $\overline{Y}'$. The two-dimensional surface point group is $D_1$.}
 \label{FigBZs}
	\end{figure}

\begin{figure*}[t]
		\centering
		\begin{center}
			\includegraphics[width = 0.98\textwidth]{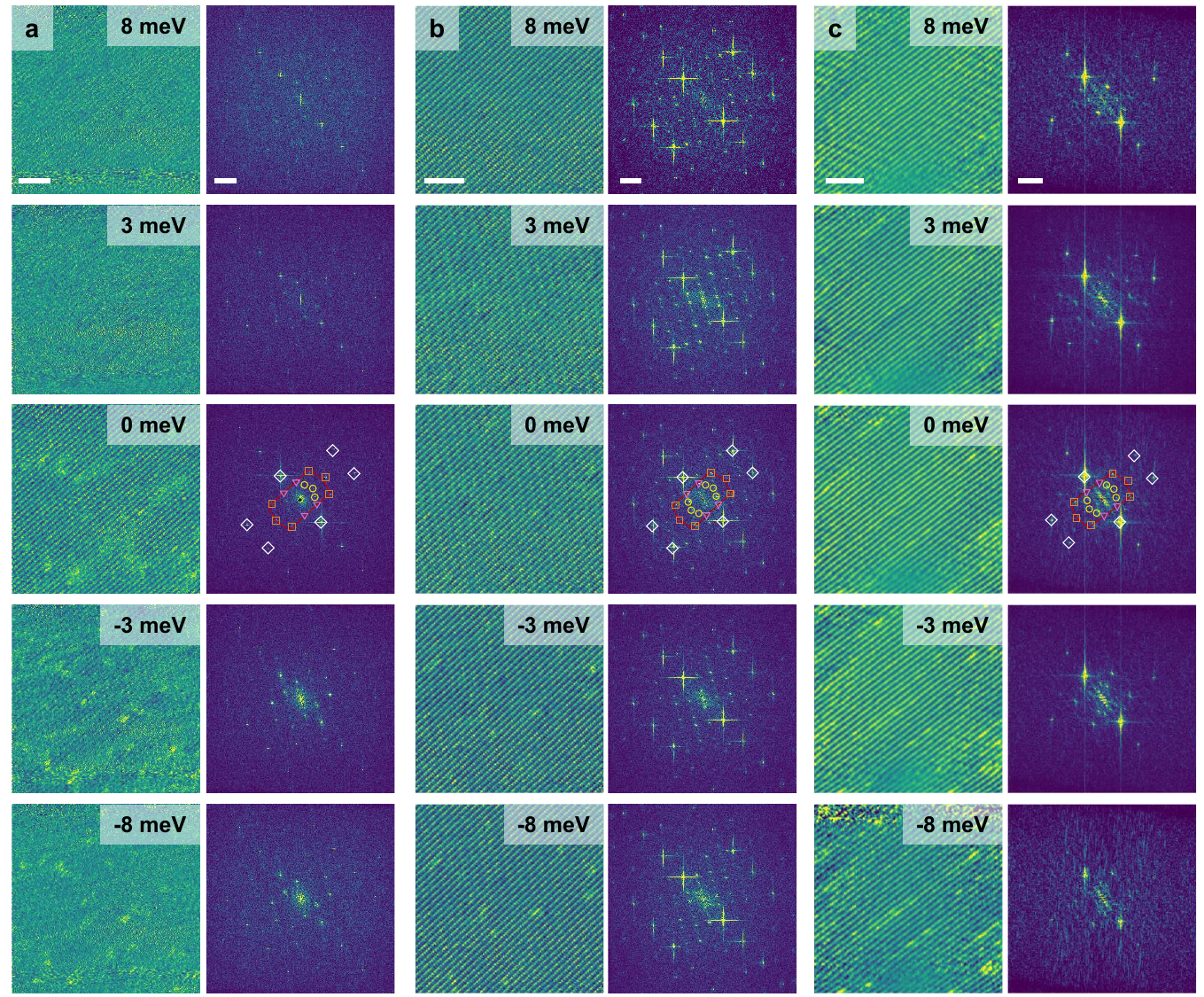}
		\end{center}
		\caption{\noindent{\bf \,| Conductance maps and their Fourier transforms.} We present representative tunneling conductance maps (left panels) and their corresponding Fourier transforms (right panels) for a few bias voltages (indicated in each panel), acquired at magnetic fields of {\bf a} 10~T, {\bf b} 15~T and {\bf c} 20~T. The full bias voltage dependence is provided as Supplementary videos.}
  \label{FigureBias}
\end{figure*}

\begin{figure*}[t]
		\centering
		\begin{center}
			\includegraphics[width = 0.45\textwidth]{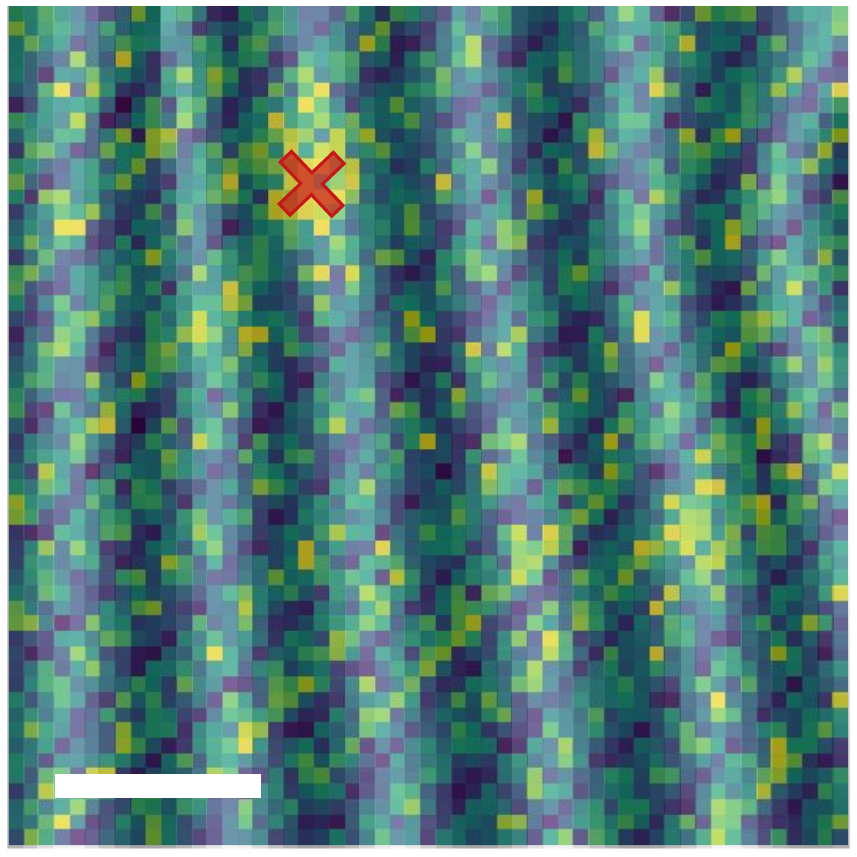}
		\end{center}
		\caption{\noindent{\bf \,| Position of the CDW with respect to defects.} The conductance map acquired at 20~T and $\approx-30$~meV, shows the $Q_{CDW2}$ oscillations highlighted by overlaid white stripes. These stripes indicate CDW oscillation crests, which tend to coincide with defect locations (red cross). The white scale bar is 2 nm long.}
  \label{FigurePhase1}
\end{figure*}

\begin{figure*}[t]
		\centering
		\begin{center}
			\includegraphics[width = 0.8\textwidth]{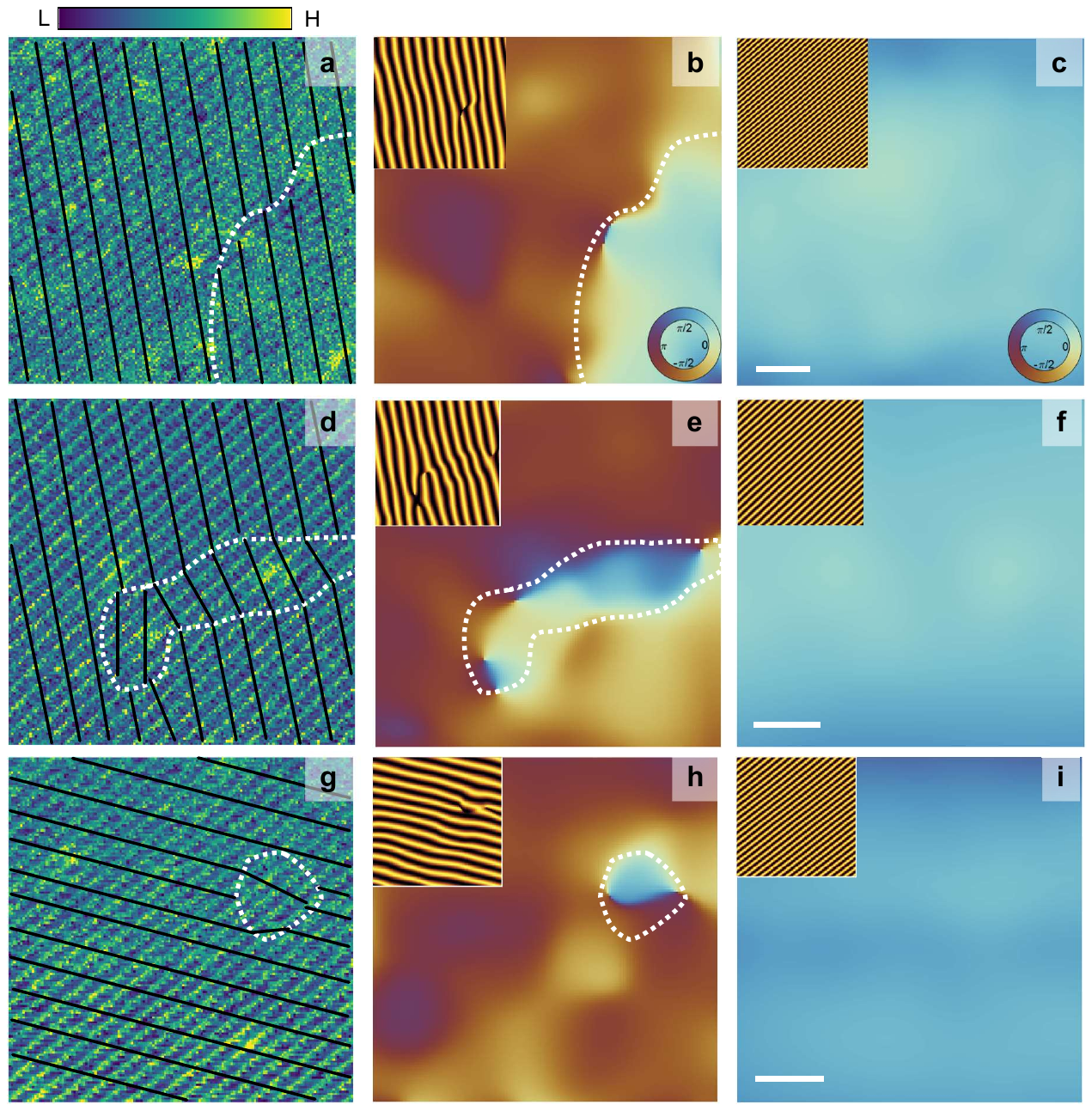}
		\end{center}
		\vskip -2 cm
		\caption{\noindent{\bf \,| Phase and amplitude of CDW and atomic oscillations.} {\bf a,d,g} We present tunneling conductance maps, corresponding to those in Fig.\,\ref{FigureCDW}{\bf a,b,c} using the color scale shown at the top of {\bf a}. Black lines delineate the crests of oscillations arising from the CDW component with the highest CDW intensity. Dashed white lines indicate CDW domain boundaries. {\bf b, e, h} The insets display the real-space images of one component of the CDW, obtained by Fourier-filtering. The main panels depict the CDW phase, $\varphi_{CDW}$, using the color scale indicated by the circle in {\bf b}. CDW domains with different phases are demarcated by dashed white lines. {\bf c, f, i} The insets show real space maps of the atomic modulation corresponding to the Te(2) chains, obtained by Fourier filtering the atomic modulations. Notably, these patterns exhibit no dislocations. Consequently, their respective phase maps, shown in the main panels, are essentially uniform. The phase is represented by the color scale shown in the circle in {\bf c}. The white bars are 3\,nm long.}
  \label{FigurePhase2}
\end{figure*}

\begin{figure*}[t]
		\centering
		\begin{center}
			\includegraphics[width = 0.75\textwidth]{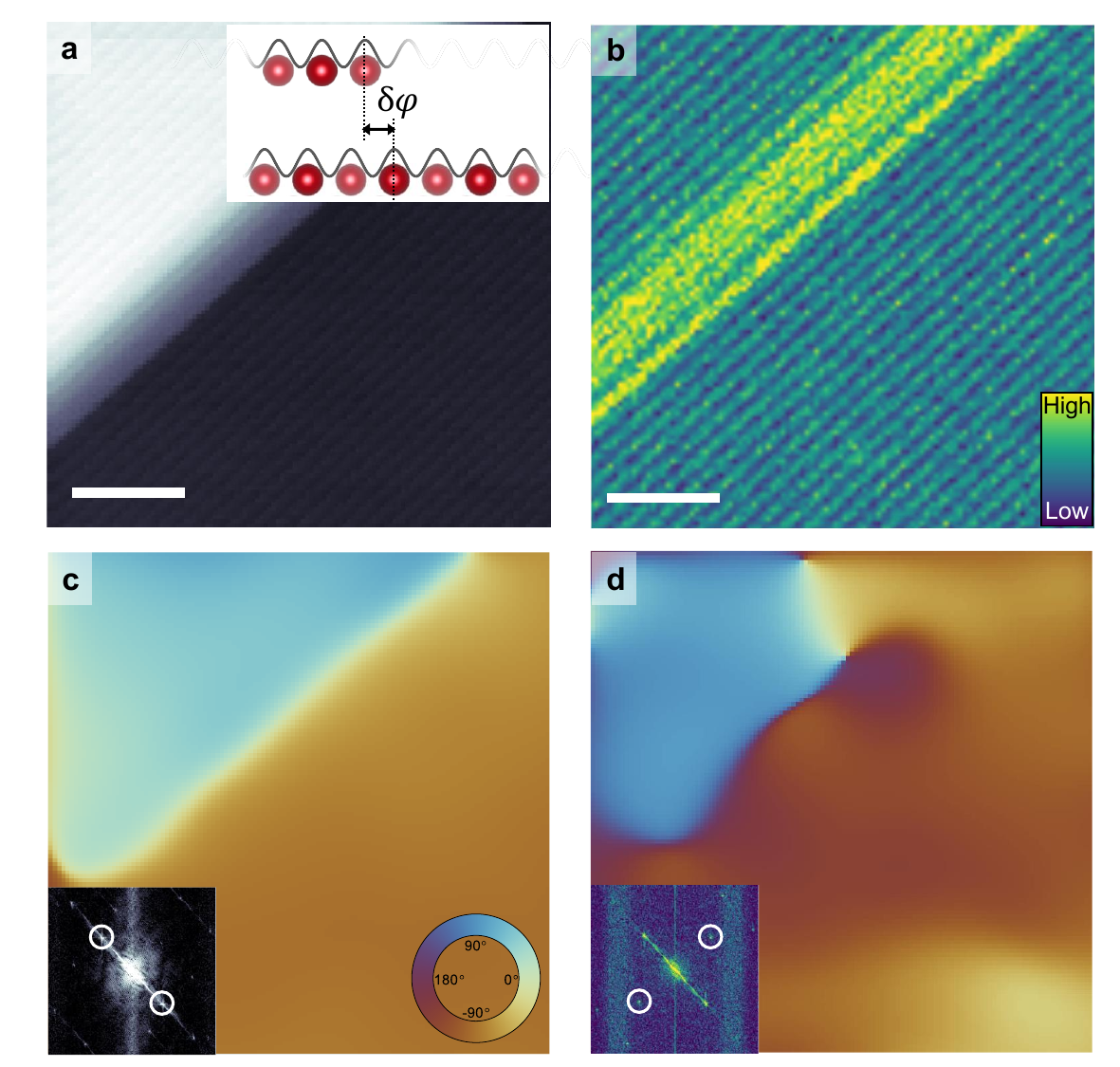}
		\end{center}
		\vskip -0.5 cm
		\caption{\noindent{\bf \,| Phase and amplitude of CDW on steps.} {\bf a} Topography map of a UTe$_2$ surface exhibiting a step. The step height corresponds to one unit cell along the direction normal to the surface. The step is oriented along the crystallographic (100) axis (the Te(2) rows). The color scale, ranging from black to white, represents a height difference of approximately 0.55 nm. The inset schematically illustrates Te(2) chains oriented perpendicular to the step. These chains are displaced by approximately one-third of the interatomic distance between Te(2) atoms on successive terraces. This corresponds to a phase shift in the atomic pattern by about $110 ^{\circ}$ in between both layers, along the direction perpendicular to the step. The white scale bar represents 5 nm. {\bf b} Tunneling conductance map, acquired at -30\,mV and at a magnetic field of 10\,T, in the same field of view as in {\bf a}.  {\bf c} Phase pattern of the atomic lattice, derived from the Fourier transform shown in the inset using the two Bragg peaks indicated by white circles. These peaks correspond to the atomic size modulation observed in {\bf a}, perpendicular to the step. The phase is represented by the color scale shown in the circle in {\bf c} and aligns with the atomic displacement expected for a perfect crystalline lattice with the surface oriented along (011) and the step along (100), as schematically depicted in the inset of {\bf a}. {\bf d} Phase pattern of the CDW, derived from the Fourier transform shown in the inset using the two Bragg peaks indicated by white circles. These peaks correspond to CDW phase modulations along the step. A phase shift of approximately $100^{\circ}$ is observed across the step. This shows that the CDW phases on each terrace are independent from each other.}
  \label{FigurePhase3}
\end{figure*}

\begin{figure*}[t]
		\centering
		\begin{center}
			\includegraphics[width = 0.98\textwidth]{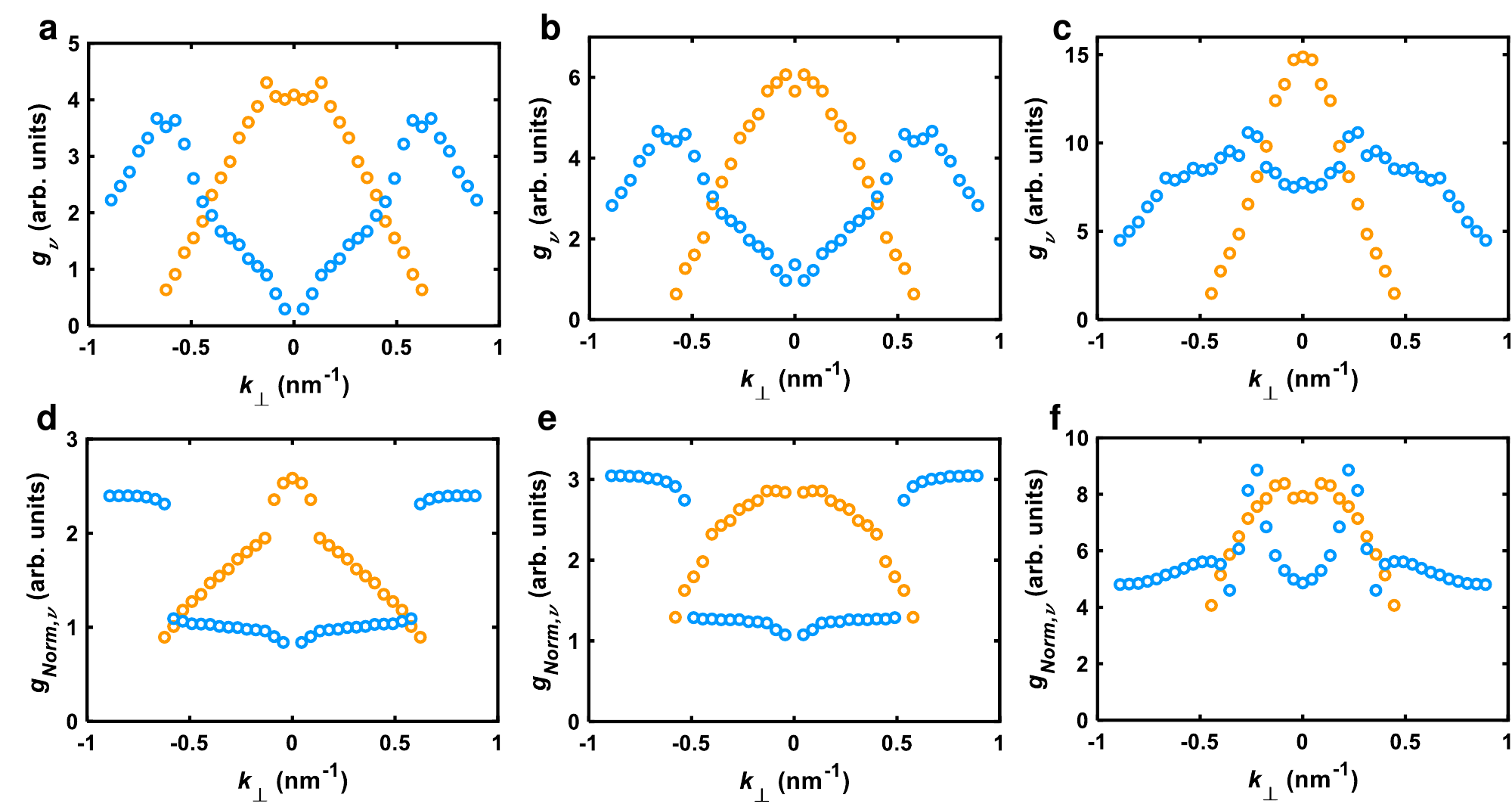}
		\end{center}
		\caption{\noindent{\bf \,| Density of states of Fermi contours at the (011) surface plane of UTe$_2$ as a function of the energy.} {\bf a-c} Calculated density of states for each band, $g_{\nu}(k_{\perp})=\int_L\frac{dL}{\vert \nabla_{\bf{k}} E\vert }$, where blue represents electron-like bands and orange hole-like bands. {\bf d-f} Calculated normalized density of states for each band, $g_{Norm,\nu}(k_{\perp},n)=\frac{1}{L}\int_L\frac{dL}{\vert \nabla_{\bf{k}} E\vert }$, with blue indicating the electron-like band and orange indicating the hole-like band. The energy is -0.1\,eV in {\bf a,d}, 0\,eV in {\bf b,e} and +0.1\,eV in {\bf c,f}.}
  \label{FigContourIntegral}
\end{figure*}

\begin{figure*}[t]
		\centering
		\begin{center}
			\includegraphics[width = 0.98\textwidth]{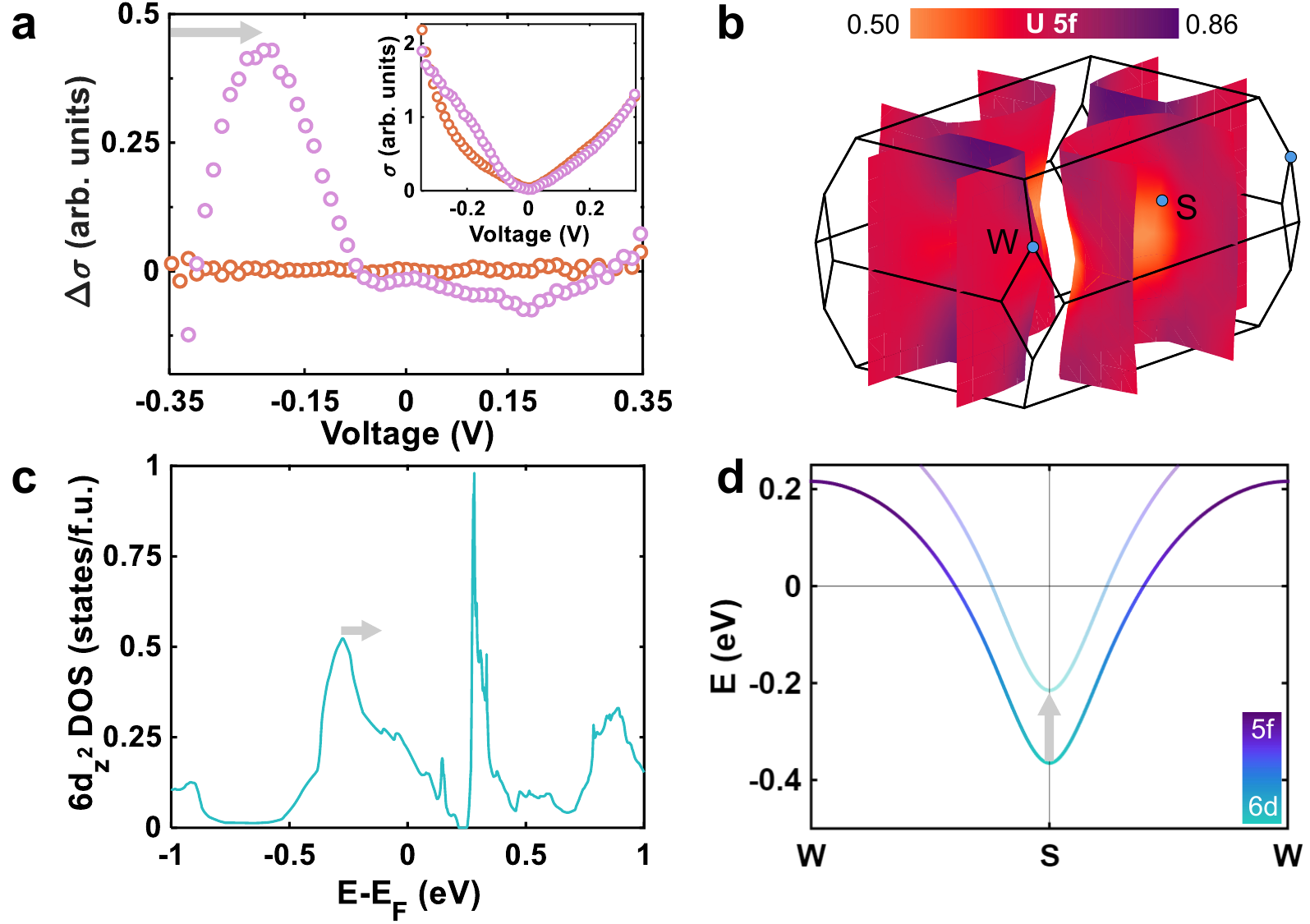}
		\end{center}
		\caption{\noindent{\bf \,| U\,5f-electron character at the Fermi surface and U\,6$\mathbf{d_{z^2}}$ character below the Fermi surface.} {\bf a} In the upper right inset we show the tunneling conductance on an atomically flat terrace (orange circles) and the tunneling conductance on a two-unit cell step (pink circles). Note the large energy range scanned in this tunneling conductance curves. In the main panel we show the same tunneling conductance curves with a subtracted background, corresponding to the average tunneling conductance on the flat terrace. {\bf b} We show as colored areas the Fermi surface of UTe$_2$ obtained from DFT. The color code provides the U\,5f character of the electronic wavefunctions, following the bar on the top. Changes are from U\,6d $+$ Te\,5p to U\,5f. {\bf c} We show in light blue the partial density of states with U\,6$\mathrm{d_{z^2}}$ character, which is dominant below the Fermi level. It has a pronounced one-dimensional character, leading to the peak at the band bottom. {\bf d} We show as a line the electron band that crosses the Fermi level and provides the $Q_{CDW1}$ nesting wavevector. The violet to blue color scale represents the strength of U\,5f and U\,6d$^2_z$ characters, following the bar on the right. In {\bf a,c,d} we schematically show by a grey arrow the possible shift in the band structure produced at the step edge, due to uncompensated charges (Extended Data Fig.\,\ref{FigSpaceFillingPolyhedra}{\bf k}). This also leads to a shift in the U\,5f valence towards U$^{4+}$ and a decrease of the Kondo hybridization at small energies (Fig.\,\ref{Steps}).}
  \label{FigBandstructure}
\end{figure*}

\begin{figure*}[t]
		\centering
		\begin{center}
			\includegraphics[width = 0.98\textwidth]{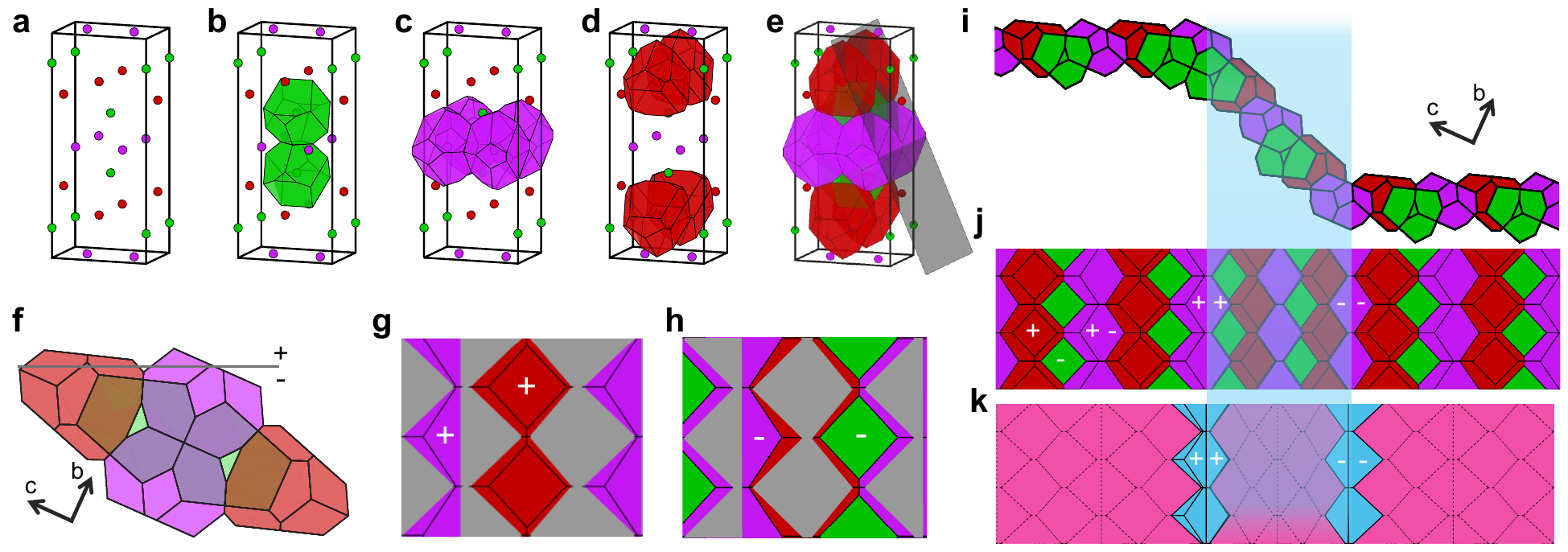}
		\end{center}
		\caption{\noindent{\bf \,| Space-filling polyhedra of UTe$_2$.} {\bf a-f} We show the UTe$_2$ unit cell by black lines and the atomic positions by colored dots. In {\bf b-e} we show, in addition, space-filling polyhedra of U (green, {\bf b}), Te(1) (violet, {\bf c}), Te(2) (red, {\bf d}) and all of them together ({\bf e}). The (011) surface plane is shown in grey in {\bf e}. See Supplementary Information, Section 5 for further details. {\bf f} Space-filling polyhedra at the surface, viewed along the a-axis. The grey line represents the surface plane, and the $+\, \mathrm{and}\, -$ signs represent the regions above and below the surface. {\bf g} Top view of the (011) surface, where we show in color (red and violet) the part of the surface-filling polyhedra above the surface plane (grey). In {\bf h} we provide the part located below the surface plane (grey). We mark by $+$ and $-$ spaces that have the same volume and exactly compensated with each other. {\bf i} Lateral view of the surface structure across a two-unit-cell step. {\bf j} View of the step from the top of the surface. On the terraces, volumes that compensate with each other are marked by $+$ and $-$ (same volumes as in {\bf g,h}), but at the step edges the volumes can no longer be compensated. {\bf k} We show the surface at a step viewed from the top. The positions at the surface where there are volumes which cannot be compensated are shown in blue.}
  \label{FigSpaceFillingPolyhedra}
\end{figure*}

\clearpage
\newpage

\section*{Supplementary Information for: Surface charge density wave in UTe$_2$}

\section{Surface and bulk Brillouin zones}
\label{Supp_BZ}

It is useful to remind that the surface Brillouin zone is the Wigner-Seitz cell formed by the projection of the bulk Brillouin zone's $\Gamma$ points onto the surface plane in reciprocal space. In general, the surface Brillouin zone differs from any cross-section of the bulk Brillouin zone defined by the surface plane, i.e. from the bulk Brillouin zone planes at any wavevector perpendicular to the surface $k_{\perp}$\,\cite{Bechstedt}. For example, in a face-centered-cubic (FCC) crystal, the surface Brillouin zone for the (100) surface is a square. The cut of the bulk Brillouin zone for the plane perpendicular to the direction $\Gamma-X$ at $k_{\perp}=|\Gamma-X|$ is also a square, but with smaller size than the surface Brillouin zone. At $k_{\perp}=0$, the cut of the bulk Brillouin zone is a parallelogram which is larger than the surface Brillouin zone.

The crystallographic relationship between the bulk and surface Brillouin zones of UTe$_2$ is illustrated in Extended Data Fig.\,1. The bulk Brillouin zone is delineated by black lines, with standard high-symmetry point labels. The $\Gamma$ points of the 14 adjacent bulk Brillouin zones are indicated by blue stars and are defined by 
the reciprocal lattice vectors ${\bm G}_S=\big(\pm\frac{1}{a},\pm\frac{1}{b},0\big)$,
$\big(\pm\frac{1}{a},0,\pm\frac{1}{c}\big)$,
$\big(0,\pm\frac{1}{b},\pm\frac{1}{c}\big)$ and
$\big(0,0,\pm\frac{1}{c}\big)$. In  Extended Data Fig.\,1{\bf b}, the bulk Brillouin zone is shown projected onto the reciprocal (011) plane. In this projection, the bulk Brillouin zone $\Gamma$ and $S$ points coincide. The surface Brillouin zone, defined as the Wigner-Seitz cell of the lattice formed by the projected bulk $\Gamma$ points (blue stars) onto the reciprocal (011) plane, is represented by the red parallelogram in Extended Data Fig.\,1{\bf b}. The high-symmetry points of the surface Brillouin zone are labelled using standard notation for a centered rectangular two-dimensional Brillouin zone, namely $\overline{\Gamma}$, $\overline{X}$, $\overline{Y}$ and $\overline{Y'}$.

It is important to note that the symmetry of the surface Brillouin zone differs from that of the surface projection of the bulk Brillouin zone. Specifically, points $L_1$ and $L_2$ are not symmetry-equivalent in the bulk, as shown in  Extended Data Fig.\,1{\bf a}. However, the corresponding points in the surface Brillouin zone, $\overline{Y'}$, which lie near the projection of $L_1$ and $L_2$ on the (011) surface, are symmetry-equivalent, as shown in Extended Data Fig.\,1{\bf b}. The two-dimensional surface point group is $D_1$.  The two-dimensional reciprocal lattice vectors of the surface Brillouin zone $\overline{{\bm G}_S}$ join the projections of the bulk $\Gamma$ points to the surface plane. For a comprehensive list of high symmetry point labels we refer to Ref.\,\cite{SETYAWAN2010299}. 

\section{CDW on different terraces in UTe$_2$}

The CDW phase, $\varphi_{CDW}$, is often determined by its interaction with defects. In UTe$_2$, we observe that, generally, maxima in the CDW modulation tend to be pinned at defects (Extended Data Fig.\,3). This leads to the formation of domains with CDWs having different phases with respect to the crystal lattice (Extended Data Fig.\,4). Furthermore, the phase is independent in different terraces. The CDW modulation locks to the defect distribution on each terrace (Extended Data Fig.\,5), demonstrating that it changes in subsequent layers.

\section{Fermi contours along k$_\perp$}
\label{Supp_FermiContours}

The surface density of states, $g_\nu(k_\perp)$, is calculated as the summation of contributions from Fermi contours within each $k_{\perp}$ plane. The tunneling conductance measured by STM is proportional to the surface density of states. However, determining the specific contributions to the tunneling conductance from individual portions of the band structure is challenging, as it depends on the tunneling matrix elements that govern tip-sample coupling. Nevertheless, we can reasonably assume that the  $k_{\perp}$ planes exhibiting significant contributions to the density of states also contribute substantially to the tunneling conductance. To quantify these contributions, we calculated the density of states for each $k_{\perp}$ plane and each band, using the expression $g_{\nu}(k_{\perp})=\int_L\frac{dL}{\vert \nabla_{\bm{k}} E\vert }$, obtaining the results shown in Extended Data Fig.\,6{\bf a-c}. We also calculated the density of states normalized by the respective contour length, $g_{Norm,\nu}(k_{\perp})=\frac{1}{L}\int_L\frac{dL}{\vert \nabla_{\bm{k}} E\vert }$, as shown in Extended Data Fig.\,6{\bf d-f}. Our analysis indicates that the largest contributions to the tunneling conductance arise from $k_{\perp}\approx 0$ for the hole-like band and $k_{\perp}\approx |\Gamma-S|$ for the electron-like band.

\section{Origin of the CDW in $\alpha-$U}
\label{Supp_Phase}

Most models for CDWs consider the cooperative interplay of lattice and electronic interactions\,\cite{doi:10.1080/00018732.2012.719674,RevModPhys.60.1129,Rossnagel_2011,Wilson01122001,Zhu04052017}. The basic mechanism is the Peierls instability of a one-dimensional lattice and is characterized by perfect nesting of Fermi points at opposite wavevectors $k_F$ and $-k_F$. This results in a dimerized lattice and an insulating ground state at low temperatures. However, perfect Fermi surface nesting is rarely observed, even in quasi one-dimensional systems\,\cite{PhysRevB.77.165135}. Many CDW materials remain metallic in the ground state. A notable example is $\alpha-$U, which displays a metallic CDW accompanied by a structural transition\,\cite{Lander01021994,PhysRevB.42.9365,PhysRevB.80.241101}.

Such electron-phonon driven CDWs can be viewed as a consequence of the dynamical susceptibility $\chi(\bm{q})$ of the electronic system presenting a peak in its real part $\chi'(\bm{q})=\operatorname{Re}(\chi(\bm{q}))$  at $\bm{q}=\boldsymbol{Q}_{CDW}$. This favors charge modulations at $\boldsymbol{Q}_{CDW}$. Nesting gives a peak on the imaginary part of the susceptibility at wave vectors joining nested Fermi surface portions at $\bm{q}=2\bm{k}_F$, $\chi''(\bm{q})=\operatorname{Im}(\chi(\bm{q}))$, which is not necessarily sufficiently strong to be reflected in $\operatorname{Re}(\chi(2\bm{k}_F))$ and lead to a CDW\,\cite{PhysRevB.77.165135,PhysRevLett.126.096401}.

The real component of the electronic susceptibility $\chi'(\bm{q})$ provides the energy change induced by a lattice perturbation with wavevector $\bm{q}$, often compensated by a change in the atomic positions\,\cite{PhysRevB.16.3127,pssb.2220560118}. We can write

\begin{equation}
\begin{aligned}
	\operatorname{Re}\left(\chi(\bm{q})\right) &= \chi'(\bm{q}) = \sum_{\bm{k}} \frac{f(\epsilon_{\bm{k}}) - f(\epsilon_{\bm{k+q}})}{\epsilon_{\bm{k}} - \epsilon_{\bm{k+q}}} \\
	\operatorname{Im}\left(\chi(\bm{q})\right) &= \chi''(\bm{q}) \quad \text{and} \quad \lim_{\omega \rightarrow 0}   \frac{\chi''(\bm{q}, \omega)}{\omega} = \sum_{\bm{k}}\delta(\epsilon_{\bm{k}}-\epsilon_F) \delta(\epsilon_{\bm{k+q}}-\epsilon_F)	
\label{equ:chi}
\end{aligned}
\end{equation}

where $f(\epsilon_{\bm{k}})$ is the Fermi-Dirac occupation function. The change in electronic energy depends only on the real part of $\chi(\bm{q})$, $\chi'(\bm{q})$.

The imaginary part of the susceptibility, or, more precisely, $\lim_{\omega \rightarrow 0}  \frac{\chi''(\bm{q})}{\omega}$ is also called the nesting function. It can be readily seen that it selects only energies at the Fermi level, and presents a peak at a given $\bm{q}$ when many wave-vectors ${\bm{k}}$ on the Fermi surface have the same Fermi energy as ${\bm{k+\bm{q}}}$, hence when $\bm{q}$ is a Fermi surface nesting vector. The nesting wavevector and the wavevector of an eventual CDW are however not directly linked. Only a peak in $\chi'(\bm{q})$ allows for the absence of screening at a given wavevector required for the establishment of a CDW. Note that $\chi'(\bm{q})$ behaves differently than $\chi''(\bm{q})$ because  $\chi'(\bm{q})$ depends on energy levels below and above the Fermi level, whereas $\chi''(\bm{q})$ is just a Fermi surface property.

It is interesting to remind that a peak in $\chi'(\bm{q})$ is found in a simple one-dimensional model with one completely filled band\,\cite{PhysRevB.77.165135}. We can consider a one-dimensional insulator, with a filled band of dispersion $\epsilon_1(\bm{k})= \cos(ka)$, and an empty valence band of dispersion $\epsilon_2(\bm{k}) = \epsilon + \cos(ka)$, with $\epsilon >2$. Then, $\chi'(\bm{q})$ displays a peak at $q = \frac{\pi}{a}$. Only the terms joining different bands contribute to Eq.\ref{equ:chi}, as these are the only ones with non-zero numerator $f(\epsilon_1(\bm{k})) - f(\epsilon_2(\bm{k+q})) = 1$. For these states, there is always a finite energy difference $\Delta E = \epsilon_1(\bm{k})- \epsilon_2(\bm{k+q})$. We see that if the energy difference becomes small, $\frac{f(\epsilon_{\bm{k}}) - f(\epsilon_{\bm{k+q}})}{\epsilon_{\bm{k}} - \epsilon_{\bm{k+q}}}$ tends towards $f'\left(\frac{(\epsilon_{\bm{k}} +\epsilon_{\bm{k+q}})}{2}\right)$. 
The derivative of $f(\epsilon)$ is peaked at $\epsilon_F$, hence the biggest contribution is for $\epsilon_{\bm{k}}$ and  $\epsilon_{\bm{k+q}}$ on different sides of $\epsilon_F$, 
and equally away from the Fermi level. 
For a larger energy difference, the ratio $\frac{f(\epsilon_{\bm{k}}) - f(\epsilon_{\bm{k+q}})}{\epsilon_{\bm{k}} - \epsilon_{\bm{k+q}}}$ remains maximum 
when $f(\epsilon_{\bm{k}}) - f(\epsilon_{\bm{k+q}})$ is largest, 
hence still when $(\epsilon_{\bm{k}} , \epsilon_{\bm{k+q}})$ remains centered around $\epsilon_F$. The condition $(\epsilon_{\bm{k}} , \epsilon_{\bm{k+q}})$ centered around $\epsilon_F$ means that both are equidistant from $\epsilon_F$ but on opposite sides (one above, the other below). Therefore, close to the Fermi level, we can have states for which $\epsilon_{\bm{k}} = \epsilon_{\bm{k+q}}$ (maximum of $f'$, the derivative of $f$) and $\frac{d\epsilon}{dk}(\bm{k}) = v_F(\bm{k})$ is exactly opposite to $\frac{d\epsilon}{dk}(\bm{k+q}) = v_F(\bm{k+q})$. When such  two levels remain centered around $\epsilon_F$, the contribution to $\chi'(\bm{q})$ is maximized\,\cite{doi:10.1126/science.252.5002.96,Pouget_2024,PhysRevB.77.165135}. Naturally, nested regions give some contribution to  $\chi'(\bm{q})$, but a peak develops most strongly  if the condition $v_F(\bm{k}) = - v_F(\bm{k+q})$ is satisfied. Hence, for the simple one-dimensional model considered, there is a peak in $\chi'(\bm{q})$ at $q = \frac{\pi}{a}$. The conditions leading to a peak in $\chi'(\bm{q})$, in particular $v_F(\bm{k}) = - v_F(\bm{k+q})$, remain for more realistic band structures\,\cite{doi:10.1126/science.252.5002.96,Pouget_2024,PhysRevB.77.165135}.

In $\alpha-$U, the CDW can be explained by considering states of equal energy above and below $E_F$, for which $v_F(\bm{k}) = - v_F(\bm{k+q})$. Indeed, nesting of narrow f-electron U derived bands were first identified as a possible source of the CDW\,\cite{PhysRevLett.81.2978}. However, the observed pressure dependence of the CDW was found to be incompatible with the near absence of pressure induced effects on the Fermi surface nesting properties\,\cite{PhysRevLett.107.136401}. It was then shown that phonon softening, observed in neutron scattering and only obtained in calculations after incorporating fully itinerant U-5f electrons, is instead responsible for the CDW\,\cite{PhysRevLett.107.136401}. Portions of the band structure not lying exactly at the Fermi level and for which $v_F(\bm{k}) = - v_F(\bm{k+q})$ produce the relevant electronic contributions to the peak in $\chi'(\bm{q})$ which leads to the CDW in $\alpha-$U\,\cite{doi:10.1126/science.252.5002.96,PhysRevLett.126.096401}.

\section{Surface CDW in UTe$_2$}

\subsection{CDW and Kondo hybridization}

The decrease in the Kondo hybridization energy scale found in our measurements at the surface reduces the relevance of Kondo screening and can favor the establishment of a peak in $\chi'(\bm{q})$. The nesting feature we observe at $\bm{Q}_{CDW1}$ connects portions of the same electron-like band (see Extended Data Table 1) and thus naturally allows for $v_F(\bm{k}) = - v_F(\bm{k+q})$. This same electron-like band has a strong U\,5f character close to the Fermi level, which decreases for lower energies towards the bottom of the band (Extended Data Fig.\,8{\bf d}). At the band bottom, approximately at $E=-0.4$ eV, the U\,6$\mathrm{d_{z^2}}$ character is much more pronounced and provides a strong peak in the density of states, reminiscent of the peak found for a one-dimensional electronic band dispersion\,\cite{Christovam24}.

It is interesting to note that the appearance of the CDW in our STM experiments is highly dependent on the bias voltage. As we show in the Extended Data Fig.\,2 and in the supplementary videos, the peaks at $\bm{Q}_{CDW1}$ and $\bm{Q}_{CDW2}$ change their intensity with energy. Generally, there is a tendency to a stronger intensity for negative energies (below $E_F$). Furthermore, impurities and defects are much better visible at negative energies. As discussed in the main text, we can associate these changes to variations in the U\,6d character of the local density of states and concomitant changes in the U\,5f valence. To further address this issue, we can calculate the cross-correlation of tunneling conductance maps, defined as \cite{Dai2014}:

\begin{equation}
CC= \frac{\sum_{x,y}[G(V_1,x,y)-\overline{G(V_1,x,y)}][G(V_2,x,y)-\overline{G(V_2,x,y)}]}{N_xN_y\sigma_{V_1}\sigma_{V_2}}
\end{equation}

at different bias voltages $V_1$ and $V_2$ ($\sigma_{V_i}$ being the standard deviation of all the conductance values found in the map, $N_{x}$, $N_{y}$ the number of points in each direction, $G(V_i,x,y)$ the tunneling conductance maps and $\overline{G(V_i,x,y)}$ the average of the tunneling conductance at the bias voltages $V_i$).

For patterns in the STM images dominated by a modulation in the atomic positions, we expect to find $CC\approx 1$ for all bias voltages. For patterns in the STM images connected to a Peierls-like CDW (presenting atomic displacements) with a gap opened in the density of states, the change in $\varphi_{CDW}$ with energy is of $\pi$ accross the gap, due to the opposed character of the band dispersion at both sides of the gap, leading to $CC\approx -1$. We find values around $CC=-0.2$ for bias voltages lying above ($+$5 mV) and below ($-$10 mV) zero bias, suggesting that the observed features are not atomic displacements, but electronic density modulations. What is more, the patterns strongly change their intensity as a function of the bias voltage, generally leading to $CC<0$ when comparing positive and negative bias voltages. This is difficult to understand within a phonon-mediated CDW with a gap opening. This reinforces the idea that spatial variations of the correlated properties of the f-electron band structure lead to the spatial variations in the local density of states, thus forming the CDW.

\subsection{Atomic relaxation at the surface}

We can discuss the influence of the surface on the electronic properties by considering the space-filling polyhedra of UTe$_2$ (Extended Data Fig.\,8)\cite{voronoi1908,bohm1996}. On the (011) surface, the space-filling polyhedra form a structure with sharp peaks and troughs (Extended Data Fig.8{\bf f}). However, the charge distribution at the surface must be smooth. This smoothing occurs due to the Smoluchowski effect, where charge spills out of peaks and fills troughs, leading to subtle variations in surface charge density\,\cite{smoluchowski1941}. To analyze this surface effect, we first identify a plane parallel to the surface (grey line in Extended Data Fig.\,8{\bf f} and grey plane in Extended Data Fig.\,8{\bf g,h}) where we can perfectly compensate volumes emerging out of the plane towards vacuum ($+$ in Extended Data Fig.\,8{\bf f,g}) with those going towards the sample ($-$ in Extended Data Fig.\,8{\bf f,h}). The charge compensation occurs on volumes that are adjacent at the surface. It leads to a modification of the space-filling polyhedra at the surface. Accordingly, the center of mass of the compensated space-filling polyhedra at the surface slightly differs from the center of mass in the bulk\cite{finnis1974}. This suggests a tendency to have atomic displacements at the surface so that the atoms inside the space-filling polyhedra that are cut at the surface are moved slighly towards the solid\cite{finnis1974}. It is important to realize that these atomic displacements exactly maintain the atomic lattice symmetry and are distinct from a surface reconstruction. Instead, these are called atomic relaxation and consist of a symmetry preserving change in the atomic lattice constant perpendicular to the surface plane. Surface relaxation occurs unavoidably in all materials, even in metals where the charge is density easily flattened at the surface\cite{Noonan88,barnett1983,methfessel1992,sun2004}.

For the case of UTe$_2$, this aspect could be addressed using DFT calculations of surface structures. However, the inclusion of correlations, crucial to lead to the CDW in UTe$_2$ at the surface is a considerable feat. Here we use instead the method of compensating space-filling polyhedra, which is based on principles of symmetry and provides comparative estimations for the influence of surface induced modifications of the band structure amonst materials with similar properties\cite{finnis1974,methfessel1992,sun2004}. We show in Table \ref{TablePolyhedra} the behavior of surface polyhedra we obtained for atomically flat surfaces of three heavy fermion systems, UTe$_2$, URu$_2$Si$_2$ and UPt$_3$ (see also Fig.\,\ref{FigSpaceHF} for a graphical representation). In URu$_2$Si$_2$ the cleaving plane is the c-axis of the tetragonal structure and we consider a surface of U atoms\cite{Schmidt10,Aynajian10,Herrera2023}. For UPt$_3$ the cleaving plane is perpendicular to the c-axis of the hexagonal structure and the surface consists of U and Pt\cite{bisset2025determiningsuperconductingorderparameter}. We find that the smallest displacements occur in UTe$_2$, suggesting that surface relaxations are smallest in this material. However, the volume which needs to be compensated on the surface of UTe$_2$ by charge motion is largest among these three heavy fermion systems, suggesting that the influence of steps can be very pronounced as compared to URu$_2$Si$_2$ and UPt$_3$.

When reaching a step the charge balance is strongly disrupted. As the surface plane changes, the volumes of the Te(2) space-filling polyhedra cannot be fully compensated, resulting in uncompensated charge density that spills out or fills each side of the step. Furthermore, on the top side of the step, adjacent U atom space-filling polyhedra touch at each other, whereas on the bottom side they are separated (Extended Data Fig.\,8{\bf k}). For two-unit-cell steps, an entire row of U atoms exhibits a different bonding environment compated to a flat surface. This perturbation significantly influences the overall band structure, as we show in the main text and in Extended Data Fig.\,7 (grey arrows).

\section{Comparison of the primitive Bragg peaks for charge order observed here with other results}
\label{Supp_Compa}

It is instructive to consider the linear combinations of $Q_{CDW1}$ and $Q_{CDW2}$ in reciprocal space, shown in Fig.\,\ref{Peaks}. We first note that all peaks observed in our experiment are linear combinations of $Q_{CDW1}$ and $Q_{CDW2}$. The previously observed CDW peaks are located at twice $Q_{CDW1}$ and $Q_{CDW2}$, the violet triangles shown in Fig.\,2 of the main text are given by $2(Q_{CDW2}-Q_{CDW1})$, the atomic scale Bragg peaks are at four times $Q_{CDW1}$ and $Q_{CDW2}$ and many other peaks can be reconstructed this way. Importantly, all peaks reported in literature regarding the CDW, see Refs.\,\cite{li2026magneticfieldtunablecommensuratemultiqcharge,KondoHybWave} are linear combinations of  $Q_{CDW1}$ and $Q_{CDW2}$.

Furthermore, careful analysis of the features expected from quasiparticle interference due to gap nodes also lead to linear combinations of $Q_{CDW1}$ and $Q_{CDW2}$. All six quasiparticle interference features due to Andreev scattering discussed in Ref.\,\cite{3kcp-w4ny} are linear combinations of $Q_{CDW1}$ and $Q_{CDW2}$. Concretely, $q_1=Q_{CDW1}$, $q_2=Q_{CDW2}$, $q_3=-2Q_{CDW1}+4Q_{CDW2}$,  $q_4=-3Q_{CDW1}+4Q_{CDW2}$,  $q_5=-2Q_{CDW1}+2Q_{CDW2}$ and  $q_6=Q_{CDW1}$. Impurity induced quasiparticle scattering provides features associated to  superconducting wavefunctions with $B_{2u}$ and  $B_{3u}$ symmetry, that are located at wavevectors that can be constructed from $Q_{CDW1}$ and $Q_{CDW2}$. Furthermore, not yet experimentally observed modes can also be obtained from $Q_{CDW1}$ and $Q_{CDW2}$ when these are far from the Brillouin zone center (light and dark green losanges in Fig.\,\ref{Peaks}{\bf a}). Only the primitive wavevectors of these modes found at $k_x=0$ marked by dark green losanges in Fig.\,\ref{Peaks} are disconnected from $Q_{CDW1}$ and $Q_{CDW2}$ and have not been observed by us either. This comparison reinforces the relevance of the Fermi surface hot spots identified here and shows that these are inherently related to symmetry properties of the lattice of UTe$_2$.

For completeness, we show a tunneling conductance map obtained at a bias voltage of 1.6~meV in Fig.\ref{CDWMod}{\bf a}. In Fig.\ref{CDWMod}{\bf b}  we show the result of Fourier filtering out the peaks corresponding to the atomic lattice. We observe a modulation of the charge density which is interspersed by variations at impurities and defects as discussed in the main text. Nearly vertical stripes along the x-axis of the real space image are visualized in Fig.\ref{CDWMod}{\bf b}.

\clearpage
\newpage

\renewcommand{\thetable}{{\bf S\arabic{table}}}
\begin{table}[]
\caption{We analyze the crystal structures of atomically flat surfaces of the heavy fermion compounds URu$_2$Si$_2$, UPt$_3$ (see also Fig.\,\ref{FigSpaceHF}). We show in the table the total volume which needs to be displaced at the surface to achieve a smooth charge distribution, the same volume divided by the size of the surface unit cell and the distance of the bulk atomic positions at the surface layer from the center of mass of the cut polyhedra (arrows and positions in Fig.\,\ref{FigSpaceHF}{\bf e,j,o}). Notice that there are two surface atoms in UTe$_2$ ((011) cleaving plane) and in UPt$_3$ ((0001) cleaving plane of the hexagonal crystal structure). In URu$_2$Si$_2$, which has a tetragonal crystal structure, we consider the cleaved surface of U with just one atom in the surface unit cell.}
\begin{center}
\begin{tabular}{ccccc}
\hline
Material & 
\begin{tabular}[c]{@{}c@{}}Total displaced\\  volume (\AA$^3$)\end{tabular} & 
\begin{tabular}[c]{@{}c@{}}Displaced \\volume per \\surface\\ unit cell (\AA)\end{tabular} &
\begin{tabular}[c]{@{}c@{}}Distance for \\ center of \\ mass 1st (\AA)\end{tabular} & 
\begin{tabular}[c]{@{}c@{}}Distance for \\ center of \\ mass 2nd (\AA)\end{tabular} 
\\ \hline 
UTe$_2$ & 15.12 & 0.159 & Te(1): 0.006 & Te(2): 0.03 \\
URu$_2$Si$_2$ & 9.22 &  0.136 & U: 0.15 & - \\
UPt$_3$ & 2.78 & 0.094 & U: 0.07 & Pt: 0.07 
\\ \hline
\end{tabular}
\end{center}
\label{TablePolyhedra}
\end{table}

\renewcommand{\thefigure}{S\arabic{figure}}
\begin{figure*}[t]
		\centering
		\begin{center}
			\includegraphics[width = 1\textwidth]{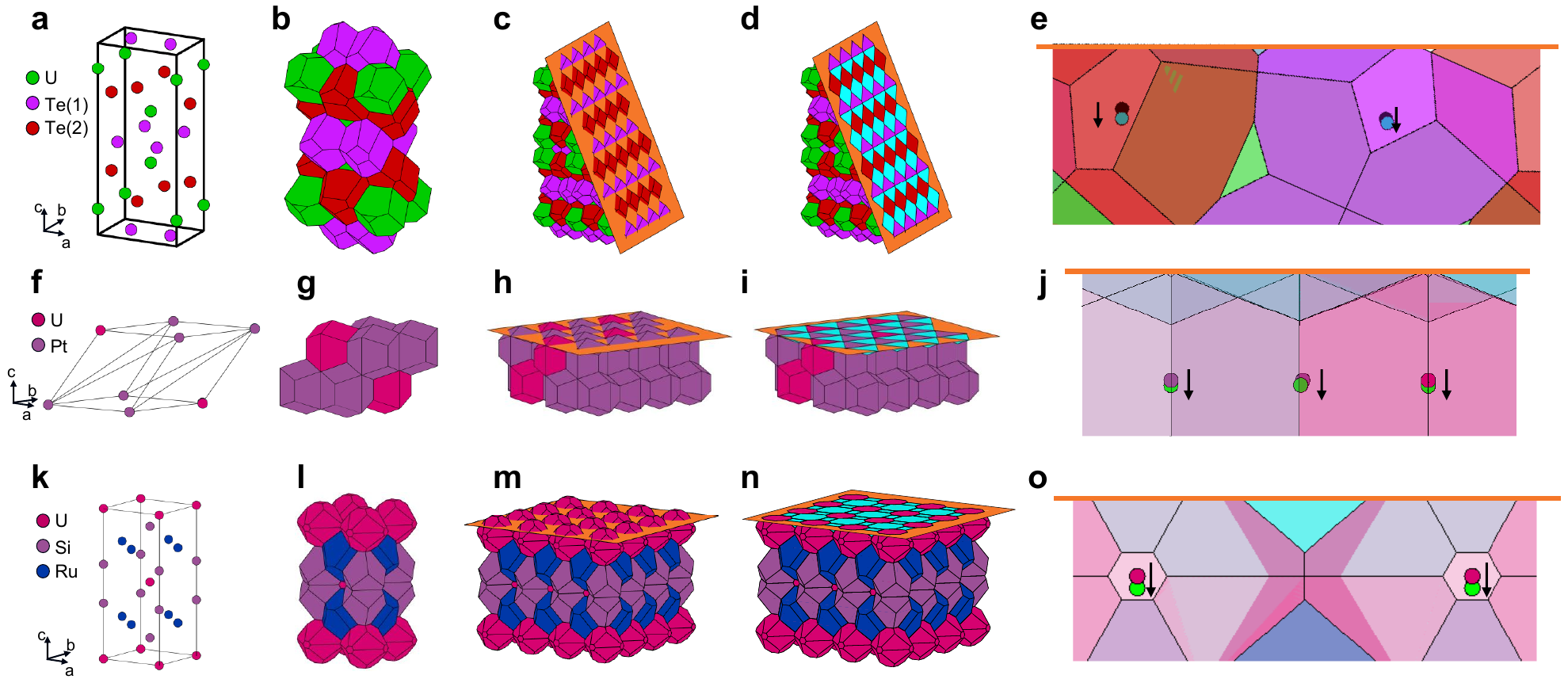}
		\end{center}
		\vskip -0cm
		\caption{\noindent{\bf \,| Surface relaxation in the heavy fermion compounds UTe$_2$, UPt$_3$ and URu$_2$Si$_2$.} {\bf a} We show the UTe$_2$ unit cell by black lines and the atomic positions by colored dots. In {\bf b} we show space-filling polyhedra of U (green), Te(1) (violet) and Te(2) (red). In {\bf c} we show the (011) surface plane in orange. In {\bf d} we show as cyan the filled volumes which lead to a flat surface. In {\bf e} we show the flat surface plane (orange line on top) and the structure of the space filling polyhedra. The black arrows point the direction of atomic relaxation, towards the bulk. Colored circles represent the center of mass of bulk (dark color) and of surface (bright color) polyhedra. {\bf f} Unit cell of UPt$_3$, with U in red and Pt in violet. {\bf g} Space filling polyhedra of bulk  UPt$_3$. The surface plane is shown in orange in {\bf h}. In {\bf i} we show a flat surface, obtained by filling the portions marked in cyan. The corresponding changes in the center of mass of space filling polyhedra at the surface is shown in {\bf j}. {\bf k-o} Equivalent figures in URu$_2$Si$_2$, with U in red, Si in violet and Ru in blue.}
  \label{FigSpaceHF}
\end{figure*}

\begin{figure*}[t]
		\centering
		\begin{center}
			\includegraphics[width = 0.98\textwidth]{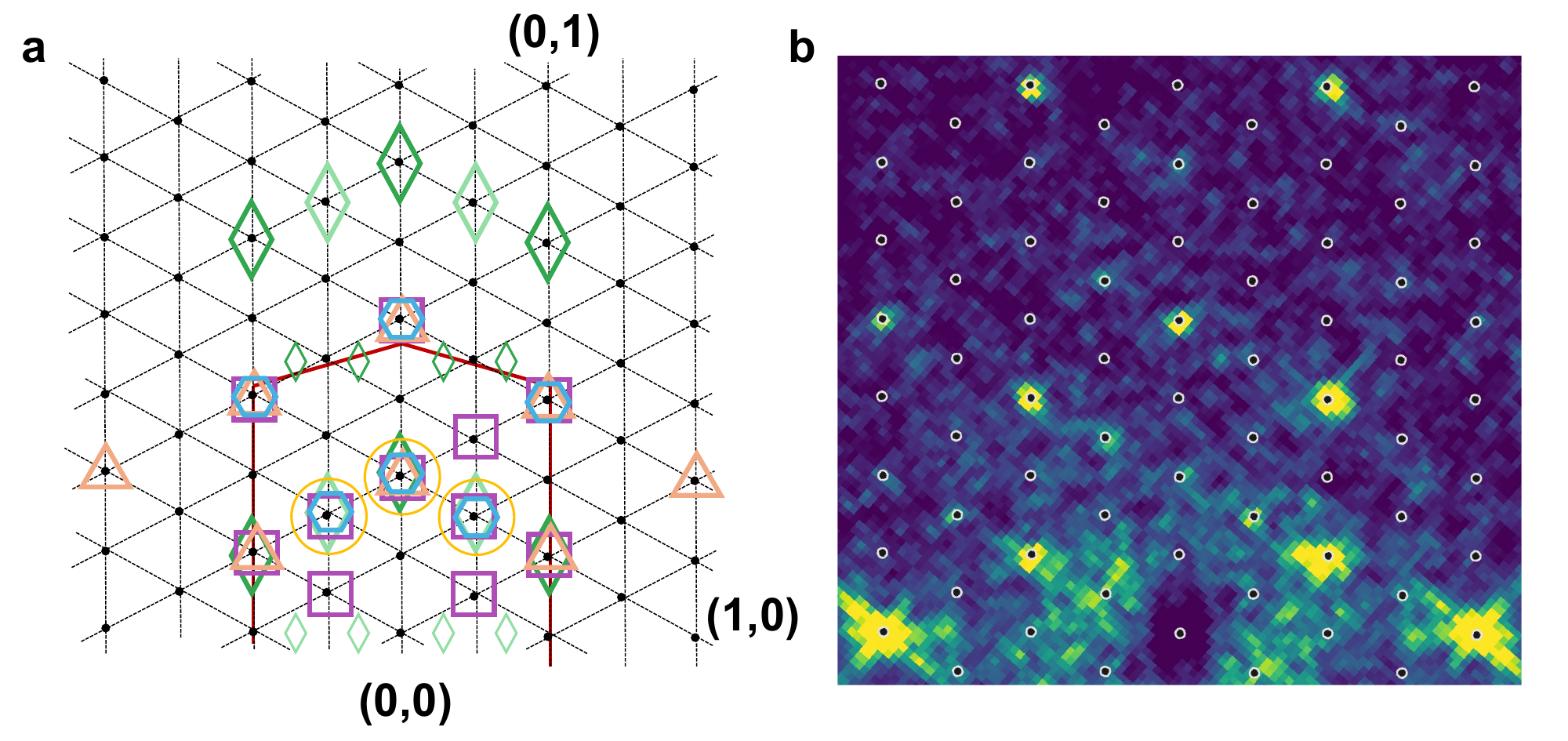}
		\end{center}
		\caption{\noindent{\bf \,| Bragg peaks in the Fourier transform of UTe$_2$.} {\bf a} We show a schematic picture of the Fourier space representing the atomic size modulations found here and those discussed by other authors. Black points provide all wavevectors obtained by a linear combination of $Q_{CDW1}$ and $Q_{CDW2}$. Violet squares provide the charge density wave peaks found in Ref.\,\cite{li2026magneticfieldtunablecommensuratemultiqcharge}, blue hexagon those discussed in Ref.\,\cite{KondoHybWave}. Furthermore, we also show the peaks expected for the superconducting order parameter discussed in Ref.\,\cite{pttl-8m71} as green losanges and those discussed in Ref.\,\cite{3kcp-w4ny} as orange triangles. Except the dark green losanges, all other peaks are linear combinations of $Q_{CDW1}$ and $Q_{CDW2}$ (yellow circles). The main surface crystalline points are shown schematically as $(0,0); (1,0)$ and $(0,1)$. {\bf b} We overlay on top of the Fourier transform of one of our tunneling conductance maps (blue to yellow) the lattice made by the primitive CDW wavevectors shown in {\bf a}.}
  \label{Peaks}
  \end{figure*}
  
\begin{figure*}[t]
		\centering
		\begin{center}
			\includegraphics[width = 0.98\textwidth]{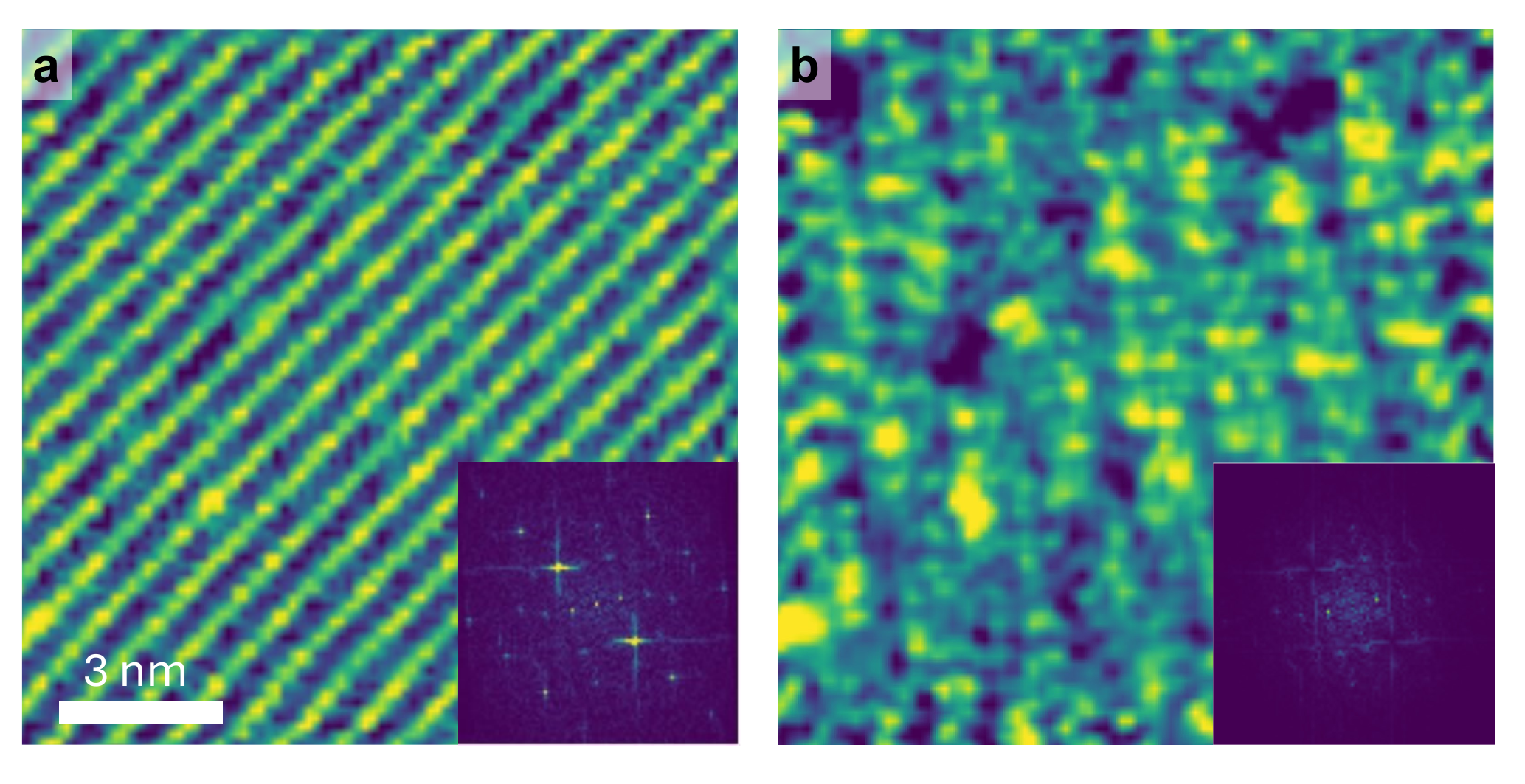}
		\end{center}
		\caption{\noindent{\bf \,| Real space atomic and CDW modulations UTe$_2$.} {\bf a} Tunneling conductance map obtained at a bias voltage of 1.7~mV, at a magnetic field of 20~T and a temperature of 80~mK. The Fourier transform is shown in the lower right inset. {\bf b} Tunneling conductance maps where we have removed atomic Bragg peaks through Fourier filtering. The Fourier transform is shown in the bottom right inset.}
  \label{CDWMod}
  \end{figure*}
 
 \clearpage
 \newpage

\bibliographystyle{naturemag}
\bibliography{bibUTe2}

@Article{Ran2019,
author={Ran, Sheng
and Liu, I-Lin
and Eo, Yun Suk
and Campbell, Daniel J.
and Neves, Paul M.
and Fuhrman, Wesley T.
and Saha, Shanta R.
and Eckberg, Christopher
and Kim, Hyunsoo
and Graf, David
and Balakirev, Fedor
and Singleton, John
and Paglione, Johnpierre
and Butch, Nicholas P.},
title={Extreme magnetic field-boosted superconductivity},
journal={Nature Physics},
year={2019},
month={Dec},
day={01},
volume={15},
number={12},
pages={1250-1254},
issn={1745-2481},
doi={10.1038/s41567-019-0670-x},
}

@Article{Aishwarya2023,
author={Aishwarya, Anuva
and May-Mann, Julian
and Raghavan, Arjun
and Nie, Laimei
and Romanelli, Marisa
and Ran, Sheng
and Saha, Shanta R.
and Paglione, Johnpierre
and Butch, Nicholas P.
and Fradkin, Eduardo
and Madhavan, Vidya},
title={Magnetic-field-sensitive charge density waves in the superconductor {UTe$_2$}},
journal={Nature},
year={2023},
month={Jun},
day={01},
volume={618},
number={7967},
pages={928-933},
issn={1476-4687},
doi={10.1038/s41586-023-06005-8},
}

@Article{Aishwarya2024,
author={Aishwarya, Anuva
and May-Mann, Julian
and Almoalem, Avior
and Ran, Sheng
and Saha, Shanta R.
and Paglione, Johnpierre
and Butch, Nicholas P.
and Fradkin, Eduardo
and Madhavan, Vidya},
title={Melting of the charge density wave by generation of pairs of topological defects in {UTe$_2$}},
journal={Nature Physics},
year={2024},
month={Jun},
day={01},
volume={20},
number={6},
pages={964-969},
issn={1745-2481},
doi={10.1038/s41567-024-02429-9},
}

@Article{Gu2023,
author={Gu, Qiangqiang
and Carroll, Joseph P.
and Wang, Shuqiu
and Ran, Sheng
and Broyles, Christopher
and Siddiquee, Hasan
and Butch, Nicholas P.
and Saha, Shanta R.
and Paglione, Johnpierre
and Davis, J. C. S{\'e}amus
and Liu, Xiaolong},
title={Detection of a pair density wave state in {UTe$_2$}},
journal={Nature},
year={2023},
month={Jun},
day={01},
volume={618},
number={7967},
pages={921-927},
issn={1476-4687},
doi={10.1038/s41586-023-05919-7},
}

@Article{LaFleur2024,
author={LaFleur, Alexander
and Li, Hong
and Frank, Corey E.
and Xu, Muxian
and Cheng, Siyu
and Wang, Ziqiang
and Butch, Nicholas P.
and Zeljkovic, Ilija},
title={Inhomogeneous high temperature melting and decoupling of charge density waves in spin-triplet superconductor {UTe$_2$}},
journal={Nature Communications},
year={2024},
month={May},
day={25},
volume={15},
number={1},
pages={4456},
issn={2041-1723},
doi={10.1038/s41467-024-48844-7},
}

@article{PhysRevB.104.205107,
title = {Thermodynamic signatures of short-range magnetic correlations in {UTe$_2$}},
author = {Willa, Kristin and Hardy, Fr\'ed\'eric and Aoki, Dai and Li, Dexin and Wiecki, Paul and Lapertot, G\'erard and Meingast, Christoph},
journal = {Phys. Rev. B},
volume = {104},
issue = {20},
pages = {205107},
numpages = {6},
year = {2021},
month = {Nov},
publisher = {American Physical Society},
doi = {10.1103/PhysRevB.104.205107},
}

@article{PhysRevB.106.L060505,
title = {$c$-axis transport in {UTe$_2$}: Evidence of three-dimensional conductivity component},
author = {Eo, Yun Suk and Liu, Shouzheng and Saha, Shanta R. and Kim, Hyunsoo and Ran, Sheng and Horn, Jarryd A. and Hodovanets, Halyna and Collini, John and Metz, Tristin and Fuhrman, Wesley T. and Nevidomskyy, Andriy H. and Denlinger, Jonathan D. and Butch, Nicholas P. and Fuhrer, Michael S. and Wray, L. Andrew and Paglione, Johnpierre},
journal = {Phys. Rev. B},
volume = {106},
issue = {6},
pages = {L060505},
numpages = {7},
year = {2022},
month = {Aug},
publisher = {American Physical Society},
doi = {10.1103/PhysRevB.106.L060505},
}

@article{FujimoriJPSJ2019,
author = {Fujimori ,Shin Ichi and Kawasaki ,Ikuto and Takeda ,Yukiharu and Yamagami ,Hiroshi and Nakamura ,Ai and Homma ,Yoshiya and Aoki ,Dai},
title = {Electronic Structure of {UTe$_2$} Studied by Photoelectron Spectroscopy},
journal = {Journal of the Physical Society of Japan},
volume = {88},
number = {10},
pages = {103701},
year = {2019},
doi = {10.7566/JPSJ.88.103701}
}

@article{Liu22,
title = {Identifying $f$-Electron Symmetries of {UTe$_2$} with {{O-edge}} Resonant Inelastic x-Ray Scattering},
author = {Liu, Shouzheng and Xu, Yishuai and Kotta, Erica C. and Miao, Lin and Ran, Sheng and Paglione, Johnpierre and Butch, Nicholas P. and Denlinger, Jonathan D. and Chuang, Yi-De and Wray, L. Andrew},
year = {2022},
month = dec,
journal = {Physical Review B},
volume = {106},
number = {24},
pages = {L241111},
publisher = {American Physical Society},
doi = {10.1103/PhysRevB.106.L241111}
}

@article{Christovam24,
	title = {Stabilization of {U 5$f^2$} Configuration in {UTe$_2$} through {U 6$d$} dimers in the Presence of {Te2} Chains},
	author = {Christovam, Denise S. and Sundermann, Martin and Marino, Andrea and Takegami, Daisuke and Falke, Johannes and Dolmantas, Paulius and Harder, Manuel and Gretarsson, Hlynur and Keimer, Bernhard and Gloskovskii, Andrei and Haverkort, Maurits W. and Elfimov, Ilya and Zwicknagl, Gertrud and Andreev, Alexander V. and Havela, Ladislav and Bordelon, Mitchell M. and Bauer, Eric D. and Rosa, Priscila F. S. and Severing, Andrea and Tjeng, Liu Hao},
	year = {2024},
	month = sep,
	journal = {Physical Review Research},
	volume = {6},
	number = {3},
	pages = {033299},
	publisher = {American Physical Society},
	doi = {10.1103/PhysRevResearch.6.033299}
}

@article{Torikachvili07,
	title = {Hydrostatic pressure study of pure and doped $\mathrm{La}_{1-x}\mathrm{R}_x\mathrm{AgSb}_2$ $(\mathrm{R}=\mathrm{Ce},\mathrm{Nd})$ charge-density-wave compounds},
	author = {Torikachvili, M. S. and Bud'ko, S. L. and Law, S. A. and Tillman, M. E. and Mun, E. D. and Canfield, P. C.},
	year = {2007},
	month = dec,
	journal = {Physical Review B},
	volume = {76},
	number = {23},
	pages = {235110},
	publisher = {American Physical Society},
	doi = {10.1103/PhysRevB.76.235110}
}

@article{Trontl,
	title = {Interplay of {{Kondo Physics}} with {{Incommensurate Charge Density Waves}} in {CeTe$_3$}},
	author = {Trontl, Vesna Miksic and Klimovskikh, Ilya I. and Kumar, Asish K. and Vyalikh, Denis V. and Louat, Alex and Cacho, Cephise and Vescovo, Elio and Vobornik, Ivana and Petrovic, Cedomir and Valla, Tonica},
	year = {2025},
	month = feb,
	primaryclass = {cond-mat},
	publisher = {arXiv},
	doi = {10.48550/arXiv.2502.04814},
	archiveprefix = {arXiv},
	journal = {arXiv:2502.04814}
}

@article{Derr06,
	title = {Valence and Magnetic Ordering in Intermediate Valence Compounds: {{TmSe}} versus {SmB$_6$}},
	shorttitle = {Valence and Magnetic Ordering in Intermediate Valence Compounds},
	author = {Derr, J and Knebel, G and Lapertot, G and Salce, B and M{\'e}asson, M-A and Flouquet, J},
	year = {2006},
	month = jan,
	journal = {Journal of Physics: Condensed Matter},
	volume = {18},
	number = {6},
	pages = {2089},
	issn = {0953-8984},
	doi = {10.1088/0953-8984/18/6/021},
	langid = {english}
}

@article{Kawakami2011,
	title = {Effects of {{Conduction Electron Correlation}} on {{Heavy-Fermion Systems}}},
	author = {Yoshida, Tsuneya and Ohashi, Takuma and Kawakami, Norio},
	year = {2011},
	month = jun,
	journal = {Journal of the Physical Society of Japan},
	volume = {80},
	number = {6},
	pages = {064710},
	publisher = {The Physical Society of Japan},
	issn = {0031-9015},
	doi = {10.1143/JPSJ.80.064710}
}

@article{Lewin_2023,
doi = {10.1088/1361-6633/acfb93},
year = {2023},
month = {oct},
publisher = {IOP Publishing},
volume = {86},
number = {11},
pages = {114501},
author = {Sylvia K Lewin and Corey E Frank and Sheng Ran and Johnpierre Paglione and Nicholas P Butch},
title = {A review of {UTe$_2$} at high magnetic fields},
journal = {Reports on Progress in Physics}
}

@article{PhysRevLett.124.076401,
  title = {Low Energy Band Structure and Symmetries of {UTe$_2$} from Angle-Resolved Photoemission Spectroscopy},
  author = {Miao, Lin and Liu, Shouzheng and Xu, Yishuai and Kotta, Erica C. and Kang, Chang-Jong and Ran, Sheng and Paglione, Johnpierre and Kotliar, Gabriel and Butch, Nicholas P. and Denlinger, Jonathan D. and Wray, L. Andrew},
  journal = {Phys. Rev. Lett.},
  volume = {124},
  issue = {7},
  pages = {076401},
  numpages = {6},
  year = {2020},
  month = {Feb},
  publisher = {American Physical Society},
  doi = {10.1103/PhysRevLett.124.076401},
}

@Article{Wu2011,
author={Wu, Tao
and Mayaffre, Hadrien
and Kr{\"a}mer, Steffen
and Horvati{\'{c}}, Mladen
and Berthier, Claude
and Hardy, W. N.
and Liang, Ruixing
and Bonn, D. A.
and Julien, Marc-Henri},
title={Magnetic-field-induced charge-stripe order in the high-temperature superconductor {YBa$_2$Cu$_3$O$_y$}},
journal={Nature},
year={2011},
month={Sep},
day={01},
volume={477},
number={7363},
pages={191-194},
issn={1476-4687},
doi={10.1038/nature10345}
}

@Article{Jiao2020,
author={Jiao, Lin
and Howard, Sean
and Ran, Sheng
and Wang, Zhenyu
and Rodriguez, Jorge Olivares
and Sigrist, Manfred
and Wang, Ziqiang
and Butch, Nicholas P.
and Madhavan, Vidya},
title={Chiral superconductivity in heavy-fermion metal {UTe$_2$}},
journal={Nature},
year={2020},
month={Mar},
day={01},
volume={579},
number={7800},
pages={523-527},
issn={1476-4687},
doi={10.1038/s41586-020-2122-2}
}

@Article{Braithwaite2019,
author={Braithwaite, D.
and Vali{\v{s}}ka, M.
and Knebel, G.
and Lapertot, G.
and Brison, J.-P.
and Pourret, A.
and Zhitomirsky, M. E.
and Flouquet, J.
and Honda, F.
and Aoki, D.},
title={Multiple superconducting phases in a nearly ferromagnetic system},
journal={Communications Physics},
year={2019},
month={Nov},
day={22},
volume={2},
number={1},
pages={147},
issn={2399-3650},
doi={10.1038/s42005-019-0248-z},
}

@misc{Githublbtuam,
  journal = {GitHub repository},
  howpublished = {\url{https://github.com/LowTemperaturesUAM}}
}

@article{fran2021,
author = {{Martín-Vega},Francisco  and Barrena,Víctor  and {Sánchez-Barquilla},Raquel  and {Fernández-Lomana},Marta  and Benito Llorens,José  and Wu,Beilun  and Fente,Antón  and Perconte Duplain,David  and Horcas,Ignacio  and López,Raquel  and Blanco,Javier  and Higuera,Juan Antonio  and {Mañas-Valero},Samuel  and Jo,Na Hyun  and Schmidt,Juan  and Canfield,Paul C.  and {Rubio-Bollinger},Gabino  and Rodrigo,José Gabriel  and Herrera,Edwin  and Guillamón,Isabel  and Suderow,Hermann },
title = {Simplified feedback control system for scanning tunneling microscopy},
journal = {Review of Scientific Instruments},
volume = {92},
number = {10},
pages = {103705},
year = {2021},
doi = {10.1063/5.0064511}
}

@article{marta2021,
author={{Fernández-Lomana},Marta  and Wu,Beilun  and {Martín-Vega},Francisco  and {Sánchez-Barquilla},Raquel  and {Álvarez-Montoya},Rafael  and Castilla,José María  and Navarrete,José  and Marijuan,Juan Ramón  and Herrera,Edwin  and Suderow,Hermann  and Guillamón,Isabel},
title={Milikelvin scanning tunneling microscope at 20/22 {T} with a graphite enabled stick-slip approach and an energy resolution 8\ $\mu$e{V}:  {A}pplication to conductance quantization at 20 {T} in single atom point contacts of {Al} and {Au} and to the charge density wave of 2{H–NbSe}$_2$},
journal = {Review of Scientific Instruments},
volume = {92},
number = {9},
pages = {093701},
year = {2021},
doi = {10.1063/5.0059394},
}

@article{suderow2011,
author = {Suderow,H.  and Guillamon,I.  and Vieira,S. },
title = {Compact very low temperature scanning tunneling microscope with mechanically driven horizontal linear positioning stage},
journal = {Review of Scientific Instruments},
volume = {82},
number = {3},
pages = {033711},
year = {2011},
doi = {10.1063/1.3567008},
}

@Article{Herrera2023,
author={Herrera, Edwin
and Guillam{\'o}n, Isabel
and Barrena, V{\'i}ctor
and Herrera, William J.
and Galvis, Jose Augusto
and Yeyati, Alfredo Levy
and Rusz, J{\'a}n
and Oppeneer, Peter M.
and Knebel, Georg
and Brison, Jean Pascal
and Flouquet, Jacques
and Aoki, Dai
and Suderow, Hermann},
title={Quantum-well states at the surface of a heavy-fermion superconductor},
journal={Nature},
year={2023},
month={Apr},
day={01},
volume={616},
number={7957},
pages={465-469},
issn={1476-4687},
doi={10.1038/s41586-023-05830-1},
}

@Article{Herrera2021,
author={Herrera, Edwin
and Barrena, V{\'i}ctor
and Guillam{\'o}n, Isabel
and Galvis, Jos{\'e} Augusto
and Herrera, William J.
and Castilla, Jos{\'e}
and Aoki, Dai
and Flouquet, Jacques
and Suderow, Hermann},
title={{1D} charge density wave in the hidden order state of {URu$_2$Si$_2$}},
journal={Communications Physics},
year={2021},
month={May},
day={14},
volume={4},
number={1},
pages={98},
issn={2399-3650},
doi={10.1038/s42005-021-00598-0}
}

@article{Rodrigo04b,
	Author = {J.G. Rodrigo and H. Suderow and S. Vieira and E. Bascones and F. Guinea},
	Journal = {J. Phys.: Condens. Matter},
	Pages = {1151},
	Title = {Superconducting nanostructures fabricated with the scanning tunnelling microscope},
	Volume = {16},
	Year = {2004},
	doi={10.1088/0953-8984/16/34/R01}
	}

@article{Horcas07,
	Author = {I. Horcas and R. Fernandez and J. M. Gomez-Rodriguez and J. Colchero and J. Gomez-Herrero and A. M. Baro},
    title = {{WSXM}: A software for scanning probe microscopy and a tool for nanotechnology},
	Journal = {Rev. Sci. Instrum.},
	Pages = {013705},
	Volume = {78},
	Year = {2007},
  doi = {10.1063/1.2432410}}

@article{PhysRevLett.125.237003,
  title = {Incommensurate Spin Fluctuations in the Spin-Triplet Superconductor Candidate {UTe$_{2}$}},
  author = {Duan, Chunruo and Sasmal, Kalyan and Maple, M. Brian and Podlesnyak, Andrey and Zhu, Jian-Xin and Si, Qimiao and Dai, Pengcheng},
  journal = {Phys. Rev. Lett.},
  volume = {125},
  issue = {23},
  pages = {237003},
  numpages = {6},
  year = {2020},
  month = {Dec},
  publisher = {American Physical Society},
  doi = {10.1103/PhysRevLett.125.237003}
}

@article{PhysRevB.104.L100409,
  title = {Low-dimensional antiferromagnetic fluctuations in the heavy-fermion paramagnetic ladder compound {UTe$_{2}$}},
  author = {Knafo, W. and Knebel, G. and Steffens, P. and Kaneko, K. and Rosuel, A. and Brison, J.-P. and Flouquet, J. and Aoki, D. and Lapertot, G. and Raymond, S.},
  journal = {Phys. Rev. B},
  volume = {104},
  issue = {10},
  pages = {L100409},
  numpages = {6},
  year = {2021},
  month = {Sep},
  publisher = {American Physical Society},
  doi = {10.1103/PhysRevB.104.L100409}
  }

@Article{Butch2022,
author={Butch, Nicholas P.
and Ran, Sheng
and Saha, Shanta R.
and Neves, Paul M.
and Zic, Mark P.
and Paglione, Johnpierre
and Gladchenko, Sergiy
and Ye, Qiang
and Rodriguez-Rivera, Jose A.},
title={Symmetry of magnetic correlations in spin-triplet superconductor {UTe$_2$}},
journal={npj Quantum Materials},
year={2022},
month={Apr},
day={05},
volume={7},
number={1},
pages={39},
issn={2397-4648},
doi={10.1038/s41535-022-00445-7},
}

@article{annurev:/content/journals/10.1146/annurev-conmatphys-031119-050711,
   author = "Agterberg, Daniel F. and Davis, J.C. Séamus and Edkins, Stephen D. and Fradkin, Eduardo and Van Harlingen, Dale J. and Kivelson, Steven A. and Lee, Patrick A. and Radzihovsky, Leo and Tranquada, John M. and Wang, Yuxuan",
   title = "The Physics of Pair-Density Waves: Cuprate Superconductors and Beyond", 
   journal= "Annual Review of Condensed Matter Physics",
   year = "2020",
   volume = "11",
   number = "Volume 11, 2020",
   pages = "231-270",
   doi = "https://doi.org/10.1146/annurev-conmatphys-031119-050711",
   publisher = "Annual Reviews",
   issn = "1947-5462",
   type = "Journal Article"
  }

@article{kengle2024absence,
    title={Absence of a bulk charge density wave signature in hard {X}-ray measurements of {UTe$_2$}},
    author={Caitlin S. Kengle and Dipanjan Chaudhuri and Xuefei Guo and Thomas A. Johnson and Simon Bettler and Wolfgang Simeth and Matthew J. Krogstad and Zahir Islam and Sheng Ran and Shanta R. Saha and Johnpierre Paglione and Nicholas P. Butch and Eduardo Fradkin and Vidya Madhavan and Peter Abbamonte},
  journal = {Phys. Rev. B},
  volume = {110},
  issue = {14},
  pages = {145101},
  year = {2024},
  month = {Oct},
  publisher = {American Physical Society},
  doi={10.1103/PhysRevB.110.145101}
}

@article{kengle2024absence2,
    title={Absence of bulk charge density wave order in the normal state of {UTe$_2$}},
    author={Caitlin S. Kengle and J. Vonka and S. Francoual and P. Abbamonte and M. Janoschek and P.F.S. Rosa and W. Simeth},
  journal = {Nat. Comm.},
  volume = {15},
  pages = {9713},
  year = {2024},
  doi={https://doi.org/10.1038/s41467-024-53739-8}
}

@article{PhysRevB.110.144507,
  title = {Absence of a bulk thermodynamic phase transition to a density wave phase in {UTe$_2$}},
  author = {Theuss, Florian and Shragai, Avi and Grissonnanche, Gael and Peralta, Luciano and Simarro, Gregorio de la Fuente and Hayes, Ian M and Saha, Shanta R and Eo, Yun Suk and Suarez, Alonso and Salinas, Andrea Capa and Pokharel, Ganesh and Wilson, Stephen D. and Butch, Nicholas P and Paglione, Johnpierre and Ramshaw, B. J.},
  journal = {Phys. Rev. B},
  volume = {110},
  issue = {14},
  pages = {144507},
  numpages = {7},
  year = {2024},
  month = {Oct},
  publisher = {American Physical Society},
  doi={10.1103/PhysRevB.110.144507}
  }

@Article{Chang2016,
author={Chang, J.
and Blackburn, E.
and Ivashko, O.
and Holmes, A. T.
and Christensen, N. B.
and H{\"u}cker, M.
and Liang, Ruixing
and Bonn, D. A.
and Hardy, W. N.
and R{\"u}tt, U.
and Zimmermann, M. v.
and Forgan, E. M.
and Hayden, S. M.},
title={Magnetic field controlled charge density wave coupling in underdoped {YBa$_2$Cu$_3$O$_{6+x}$}},
journal={Nature Communications},
year={2016},
month={May},
day={05},
volume={7},
number={1},
pages={11494},
issn={2041-1723},
doi={10.1038/ncomms11494}
}

@article{
doi:10.1126/science.1242996,
author = {R. Comin  and A. Frano  and M. M. Yee  and Y. Yoshida  and H. Eisaki  and E. Schierle  and E. Weschke  and R. Sutarto  and F. He  and A. Soumyanarayanan  and Yang He  and M. Le Tacon  and I. S. Elfimov  and Jennifer E. Hoffman  and G. A. Sawatzky  and B. Keimer  and A. Damascelli },
title = {Charge Order Driven by {Fermi}-Arc Instability in {Bi$_2$Sr$_{2-x}$La$_x$CuO$_{6+\delta}$}},
journal = {Science},
volume = {343},
number = {6169},
pages = {390-392},
year = {2014},
doi = {10.1126/science.1242996}}

@article{doi:10.1080/00018732.2012.719674,
author = {Pierre Monceau},
title = {Electronic crystals: an experimental overview},
journal = {Advances in Physics},
volume = {61},
number = {4},
pages = {325--581},
year = {2012},
publisher = {Taylor \& Francis},
doi = {10.1080/00018732.2012.719674}
}

@article{liu2024density,
    title={Density functional theory based investigation of heavy fermion band candidates in triplet superconductor {UTe$_2$}},
    author={Shouzheng Liu and L. Andrew Wray},
    year={2024},
    archivePrefix={arXiv},
    primaryClass={cond-mat.supr-con},
    journal = {arXiv:2410.03840}
}

@article{shimizu2025,
    title={Electronic structure of {UTe$_2$} under pressure},
    author={Makoto Shimizu and Youichi Yanase},
journal = {Journal of the Physical Society of Japan},
volume = {94},
pages = {124708},
year = {2025}
}

@ARTICLE{Haga2022,
  title    = {Effect of uranium deficiency on normal and superconducting properties in unconventional superconductor {UTe$_2$}},
  author   = {Haga, Y and Opletal, P and Tokiwa, Y and Yamamoto, E and
              Tokunaga, Y and Kambe, S and Sakai, H},
  journal  = {J. Phys.: Condens. Matter},
  volume   =  {34},
  number   =  {17},
  year     =  {2022},
  doi = {10.1088/1361-648X/ac5201},
  pages = {175601}
}

@ARTICLE{Sakai2022,
  title     = "Single crystal growth of superconducting {UTe$_{2}$} by molten
               salt flux method",
  author    = "Sakai, H and Opletal, P and Tokiwa, Y and Yamamoto, E and
               Tokunaga, Y and Kambe, S and Haga, Y",
  journal   = "Phys. Rev. Mater.",
  publisher = "American Physical Society",
  volume    =  6,
  number    =  7,
  pages     = "073401",
  month     =  jul,
  year      =  2022,
  doi       = {10.1103/PhysRevMaterials.6.073401}
}

@article{Koepernik1999,
  title = {Full-potential nonorthogonal local-orbital minimum-basis band-structure scheme},
  author = {Koepernik, Klaus and Eschrig, Helmut},
  journal = {Phys. Rev. B},
  volume = {59},
  issue = {3},
  pages = {1743--1757},
  numpages = {0},
  year = {1999},
  month = {Jan},
  publisher = {American Physical Society},
  doi = {10.1103/PhysRevB.59.1743},
  xurl = {https://link.aps.org/doi/10.1103/PhysRevB.59.1743}
}

@article{Ylvisaker2009,
  title     = "Anisotropy and magnetism in the $\text{LSDA}+\text{U}$ method",
  author    = "Ylvisaker, Erik R and Pickett, Warren E and Koepernik, Klaus",
  journal   = "Phys. Rev. B Condens. Matter",
  publisher = "American Physical Society",
  volume    =  79,
  number    =  3,
  pages     = "035103",
  month     =  jan,
  year      =  2009,
  doi       = {10.1103/PhysRevB.79.035103},
}

@article{Perdew1996,
  title = {Generalized Gradient Approximation Made Simple},
  author = {Perdew, John P. and Burke, Kieron and Ernzerhof, Matthias},
  journal = {Phys. Rev. Lett.},
  volume = {77},
  issue = {18},
  pages = {3865--3868},
  numpages = {0},
  year = {1996},
  month = {Oct},
  publisher = {American Physical Society},
  doi = {10.1103/PhysRevLett.77.3865},
  xurl = {https://link.aps.org/doi/10.1103/PhysRevLett.77.3865}
}

@article{Ishizuka2019,
  title = {Insulator-Metal Transition and Topological Superconductivity in {UTe$_2$} from a First-Principles Calculation},
  author = {Ishizuka, Jun and Sumita, Shuntaro and Daido, Akito and Yanase, Youichi},
  journal = {Phys. Rev. Lett.},
  volume = {123},
  issue = {21},
  pages = {217001},
  numpages = {6},
  year = {2019},
  month = {Nov},
  publisher = {American Physical Society},
  doi = {10.1103/PhysRevLett.123.217001}
}

@INPROCEEDINGS{Harima2020,
  title     = "How to Obtain {Fermi} Surfaces of {UTe$_2$}",
  booktitle = "Proceedings of {J-Physics} 2019: International Conference on
               Multipole Physics and Related Phenomena",
  author    = "Harima, Hisatomo",
  publisher = "Journal of the Physical Society of Japan",
  volume    =  29,
  series    = "JPS Conference Proceedings",
  month     =  feb,
  year      =  2020,
 doi        = {10.7566/JPSCP.29.011006}
}

@article{Aoki_dHvA2022,
author = {Aoki ,Dai and Sakai ,Hironori and Opletal ,Petr and Tokiwa ,Yoshifumi and Ishizuka ,Jun and Yanase ,Youichi and Harima ,Hisatomo and Nakamura ,Ai and Li ,Dexin and Homma ,Yoshiya and Shimizu ,Yusei and Knebel ,Georg and Flouquet ,Jacques and Haga ,Yoshinori},
title = {First Observation of the de {Haas}-van {Alphen} Effect and {Fermi} Surfaces in the Unconventional Superconductor {UTe$_2$}},
journal = {Journal of the Physical Society of Japan},
volume = {91},
number = {8},
pages = {083704},
year = {2022},
doi = {10.7566/JPSJ.91.083704}
}

@article{Aoki_dHvA2023,
author = {Aoki ,Dai and Sheikin ,Ilya and McCollam ,Alix and Ishizuka ,Jun and Yanase ,Youichi and Lapertot ,Gerard and Flouquet ,Jacques and Knebel ,Georg},
title = {de {Haas}-van {Alphen} Oscillations for the Field Along c-axis in {UTe$_2$}},
journal = {Journal of the Physical Society of Japan},
volume = {92},
number = {6},
pages = {065002},
year = {2023},
doi = {10.7566/JPSJ.92.065002}
}

@article{Broyles2023,
  title = {Revealing a {3D} {Fermi} Surface Pocket and Electron-Hole Tunneling in {UTe$_2$} with Quantum Oscillations},
  author = {Broyles, Christopher and Rehfuss, Zack and Siddiquee, Hasan and Zhu, Jiahui Althena and Zheng, Kaiwen and Nikolo, Martin and Graf, David and Singleton, John and Ran, Sheng},
  journal = {Phys. Rev. Lett.},
  volume = {131},
  issue = {3},
  pages = {036501},
  numpages = {7},
  year = {2023},
  month = {Jul},
  publisher = {American Physical Society},
  doi = {10.1103/PhysRevLett.131.036501}
}

@Article{Eaton2024,
author={Eaton, A. G.
and Weinberger, T. I.
and Popiel, N. J. M.
and Wu, Z.
and Hickey, A. J.
and Cabala, A.
and Posp{\'i}{\v{s}}il, J.
and Prokle{\v{s}}ka, J.
and Haidamak, T.
and Bastien, G.
and Opletal, P.
and Sakai, H.
and Haga, Y.
and Nowell, R.
and Benjamin, S. M.
and Sechovsk{\'y}, V.
and Lonzarich, G. G.
and Grosche, F. M.
and Vali{\v{s}}ka, M.},
title={Quasi-{2D} {Fermi} surface in the anomalous superconductor {UTe$_2$}},
journal={Nature Communications},
year={2024},
month={Jan},
day={03},
volume={15},
number={1},
pages={223},
issn={2041-1723},
doi={10.1038/s41467-023-44110-4}
}

@article{AokiReview2022,
	title = {Unconventional Superconductivity in {{UTe$_2$}}},
	author = {Aoki, D and Brison, J-P and Flouquet, J and Ishida, K and Knebel, G and Tokunaga, Y and Yanase, Y},
	year = {2022},
	month = apr,
	journal = {Journal of Physics: Condensed Matter},
	volume = {34},
	number = {24},
	pages = {243002},
	publisher = {IOP Publishing},
	issn = {0953-8984},
	doi = {10.1088/1361-648X/ac5863},
	langid = {english}
}

@article{Taillefer88,
	title = {Heavy-Fermion Quasiparticles in {UPt$_3$}},
	author = {Taillefer, L. and Lonzarich, G. G.},
	year = {1988},
	month = apr,
	journal = {Physical Review Letters},
	volume = {60},
	number = {15},
	pages = {1570},
	publisher = {American Physical Society},
	doi = {10.1103/PhysRevLett.60.1570}
}

@article{Weinberger2024,
  title = {Quantum Interference between Quasi-{2D} {Fermi} Surface Sheets in {UTe$_2$}},
  author = {Weinberger, T. I. and Wu, Z. and Graf, D. E. and Skourski, Y. and Cabala, A. and Posp\'{\i}\ifmmode \check{s}\else \v{s}\fi{}il, J. and Prokle\ifmmode \check{s}\else \v{s}\fi{}ka, J. and Haidamak, T. and Bastien, G. and Sechovsk\'y, V. and Lonzarich, G. G. and Vali\ifmmode \check{s}\else \v{s}\fi{}ka, M. and Grosche, F. M. and Eaton, A. G.},
  journal = {Phys. Rev. Lett.},
  volume = {132},
  issue = {26},
  pages = {266503},
  numpages = {8},
  year = {2024},
  month = {Jun},
  publisher = {American Physical Society},
  doi = {10.1103/PhysRevLett.132.266503}
}

@article{Miao2020,
  title = {Low Energy Band Structure and Symmetries of {UTe$_2$} from Angle-Resolved Photoemission Spectroscopy},
  author = {Miao, Lin and Liu, Shouzheng and Xu, Yishuai and Kotta, Erica C. and Kang, Chang-Jong and Ran, Sheng and Paglione, Johnpierre and Kotliar, Gabriel and Butch, Nicholas P. and Denlinger, Jonathan D. and Wray, L. Andrew},
  journal = {Phys. Rev. Lett.},
  volume = {124},
  issue = {7},
  pages = {076401},
  numpages = {6},
  year = {2020},
  month = {Feb},
  publisher = {American Physical Society},
  doi = {10.1103/PhysRevLett.124.076401}
}

@book{Bechstedt,
    title = {Principles of {{Surface Physics}}},
    author = {Bechstedt, Friedhelm},
    year = {2003},
    series = {Advanced {{Texts}} in {{Physics}}},
    publisher = {Springer},
    address = {Berlin, Heidelberg},
    doi = {10.1007/978-3-642-55466-7},
    copyright = {http://www.springer.com/tdm},
    isbn = {978-3-642-62458-2 978-3-642-55466-7},
}

@article{
doi:10.1126/science.252.5002.96,
author = {M.-H. Whangbo  and E. Canadell  and P. Foury  and J. -P. Pouget },
title = {Hidden {Fermi} Surface Nesting and Charge Density Wave Instability in Low-Dimensional Metals},
journal = {Science},
volume = {252},
number = {5002},
pages = {96-98},
year = {1991},
doi = {10.1126/science.252.5002.96}
}

@article{PhysRevB.19.6615,
  title = {Valence state at the surface of rare-earth metals},
  author = {Johansson, B\"orje},
  journal = {Phys. Rev. B},
  volume = {19},
  issue = {12},
  pages = {6615--6619},
  numpages = {0},
  year = {1979},
  month = {Jun},
  publisher = {American Physical Society},
  doi = {10.1103/PhysRevB.19.6615}
}

@article{SETYAWAN2010299,
title = {High-throughput electronic band structure calculations: Challenges and tools},
journal = {Computational Materials Science},
volume = {49},
number = {2},
pages = {299-312},
year = {2010},
issn = {0927-0256},
doi = {https://doi.org/10.1016/j.commatsci.2010.05.010},
author = {Wahyu Setyawan and Stefano Curtarolo}
}

@Article{MatsumuraJPSJ2023,
  author   = {Matsumura ,Hiroki and Fujibayashi ,Hiroki and Kinjo ,Katsuki and Kitagawa ,Shunsaku and Ishida ,Kenji and Tokunaga ,Yo and Sakai ,Hironori and Kambe ,Shinsaku and Nakamura ,Ai and Shimizu ,Yusei and Homma ,Yoshiya and Li ,Dexin and Honda ,Fuminori and Aoki ,Dai},
  journal  = {Journal of the Physical Society of Japan},
  title    = {Large Reduction in the a-axis {Knight} Shift on {UTe$_2$} with {$T_c =$} 2.1 {K}},
  year     = {2023},
  number   = {6},
  pages    = {063701},
  volume   = {92},
   doi      = {10.7566/JPSJ.92.063701},
}

@Article{SakaiPRL2023,
  author    = {Sakai, H. and Tokiwa, Y. and Opletal, P. and Kimata, M. and Awaji, S. and Sasaki, T. and Aoki, D. and Kambe, S. and Tokunaga, Y. and Haga, Y.},
  journal   = {Phys. Rev. Lett.},
  title     = {Field Induced Multiple Superconducting Phases in {UTe$_2$} along Hard Magnetic Axis},
  year      = {2023},
  month     = {May},
  pages     = {196002},
  volume    = {130},
  doi       = {10.1103/PhysRevLett.130.196002},
  issue     = {19},
  numpages  = {6},
  publisher = {American Physical Society},
}

@Article{KnebelJPSJ2019,
  author  = {Knebel ,Georg and Knafo ,William and Pourret ,Alexandre and Niu ,Qun and Vali\v{s}ka ,Michal and Braithwaite ,Daniel and Lapertot ,Gérard and Nardone ,Marc and Zitouni ,Abdelaziz and Mishra ,Sanu and Sheikin ,Ilya and Seyfarth ,Gabriel and Brison ,Jean-Pascal and Aoki ,Dai and Flouquet ,Jacques},
  journal = {Journal of the Physical Society of Japan},
  title   = {Field-Reentrant Superconductivity Close to a Metamagnetic Transition in the Heavy-Fermion Superconductor {UTe$_2$}},
  year    = {2019},
  number  = {6},
  pages   = {063707},
  volume  = {88},
  doi     = {10.7566/JPSJ.88.063707},
}

@Article{RosuelPRX2023,
  author    = {Rosuel, A. and Marcenat, C. and Knebel, G. and Klein, T. and Pourret, A. and Marquardt, N. and Niu, Q. and Rousseau, S. and Demuer, A. and Seyfarth, G. and Lapertot, G. and Aoki, D. and Braithwaite, D. and Flouquet, J. and Brison, J. P.},
  journal   = {Phys. Rev. X},
  title     = {Field-Induced Tuning of the Pairing State in a Superconductor},
  year      = {2023},
  month     = {Feb},
  pages     = {011022},
  volume    = {13},
  doi       = {10.1103/PhysRevX.13.011022},
  issue     = {1},
  numpages  = {26},
  publisher = {American Physical Society},
}

@Article{KinjoPRB2023,
  author    = {Kinjo, K. and Fujibayashi, H. and Kitagawa, S. and Ishida, K. and Tokunaga, Y. and Sakai, H. and Kambe, S. and Nakamura, A. and Shimizu, Y. and Homma, Y. and Li, D. X. and Honda, F. and Aoki, D. and Hiraki, K. and Kimata, M. and Sasaki, T.},
  journal   = {Phys. Rev. B},
  title     = {Change of superconducting character in {UTe$_2$} induced by magnetic field},
  year      = {2023},
  month     = {Feb},
  pages     = {L060502},
  volume    = {107},
  doi       = {10.1103/PhysRevB.107.L060502},
  issue     = {6},
  numpages  = {5},
  publisher = {American Physical Society},
}

@Article{AokiJPSJ2020,
  author  = {Aoki ,Dai and Honda ,Fuminori and Knebel ,Georg and Braithwaite ,Daniel and Nakamura ,Ai and Li ,DeXin and Homma ,Yoshiya and Shimizu ,Yusei and Sato ,Yoshiki J. and Brison ,Jean-Pascal and Flouquet ,Jacques},
  journal = {Journal of the Physical Society of Japan},
  title   = {Multiple Superconducting Phases and Unusual Enhancement of the Upper Critical Field in {UTe$_2$}},
  year    = {2020},
  number  = {5},
  pages   = {053705},
  volume  = {89},
   doi     = {10.7566/JPSJ.89.053705},
}

@Article{LinNPJQuMat2020,
  author   = {Lin, Wen-Chen and Campbell, Daniel J. and Ran, Sheng and Liu, I.-Lin and Kim, Hyunsoo and Nevidomskyy, Andriy H. and Graf, David and Butch, Nicholas P. and Paglione, Johnpierre},
  journal  = {npj Quantum Materials},
  title    = {Tuning magnetic confinement of spin-triplet superconductivity},
  year     = {2020},
  issn     = {2397-4648},
  number   = {1},
  pages    = {68},
  volume   = {5},
  doi      = {10.1038/s41535-020-00270-w},
  refid    = {Lin2020},
}

@Article{ThomasPRB2021,
  author    = {S. M. Thomas and C. Stevens and F. B. Santos and S. S. Fender and E. D. Bauer and F. Ronning and J. D. Thompson and A. Huxley and P. F. S. Rosa},
  journal   = {Physical Review B},
  title     = {Spatially inhomogeneous superconductivity in {UTe$_2$}},
  year      = {2021},
  month     = {dec},
  number    = {22},
  pages     = {224501},
  volume    = {104},
  doi       = {10.1103/physrevb.104.224501},
  publisher = {American Physical Society ({APS})},
}

@article{PhysRevB.77.165135,
  title = {Fermi surface nesting and the origin of charge density waves in metals},
  author = {Johannes, M. D. and Mazin, I. I.},
  journal = {Phys. Rev. B},
  volume = {77},
  issue = {16},
  pages = {165135},
  numpages = {8},
  year = {2008},
  month = {Apr},
  publisher = {American Physical Society},
  doi = {10.1103/PhysRevB.77.165135},
}

@article{RevModPhys.60.1129,
  title = {The dynamics of charge-density waves},
  author = {Gr\"uner, G.},
  journal = {Rev. Mod. Phys.},
  volume = {60},
  issue = {4},
  pages = {1129--1181},
  numpages = {0},
  year = {1988},
  month = {Oct},
  publisher = {American Physical Society},
  doi = {10.1103/RevModPhys.60.1129},
}

@article{Rossnagel_2011,
doi = {10.1088/0953-8984/23/21/213001},
year = {2011},
month = {may},
publisher = {},
volume = {23},
number = {21},
pages = {213001},
author = {K Rossnagel},
title = {On the origin of charge-density waves in select layered transition-metal dichalcogenides},
journal = {Journal of Physics: Condensed Matter}
}

@article{Wilson01122001,
author = {J. A. Wilson, F. J. Di Salvo and S. Mahajan},
title = {Charge-density waves and superlattices in the metallic layered transition metal dichalcogenides},
journal = {Advances in Physics},
volume = {50},
number = {8},
pages = {1171--1248},
year = {2001},
publisher = {Taylor \& Francis},
doi = {10.1080/00018730110102718}
}

@article{Zhu04052017,
author = {Xuetao Zhu, Jiandong Guo, Jiandi Zhang and E. W. Plummer},
title = {Misconceptions associated with the origin of charge density waves},
journal = {Advances in Physics: X},
volume = {2},
number = {3},
pages = {622--640},
year = {2017},
publisher = {Taylor \& Francis},
doi = {10.1080/23746149.2017.1343098}
}

@article{PhysRevLett.81.2978,
  title = {Theoretical Aspects of the Charge Density Wave in Uranium},
  author = {Fast, Lars and Eriksson, Olle and Johansson, B\"orje and Wills, J. M. and Straub, G. and Roeder, H. and Nordstr\"om, Lars},
  journal = {Phys. Rev. Lett.},
  volume = {81},
  issue = {14},
  pages = {2978--2981},
  numpages = {0},
  year = {1998},
  month = {Oct},
  publisher = {American Physical Society},
  doi = {10.1103/PhysRevLett.81.2978}
}

@article{PhysRevB.42.9365,
  title = {Neutron-diffraction study of the charge-density wave in $\alpha$-uranium},
  author = {Marmeggi, J. C. and Lander, G. H. and van Smaalen, S. and Br\"uckel, T. and Zeyen, C. M. E.},
  journal = {Phys. Rev. B},
  volume = {42},
  issue = {15},
  pages = {9365--9376},
  numpages = {0},
  year = {1990},
  month = {Nov},
  publisher = {American Physical Society},
  doi = {10.1103/PhysRevB.42.9365}
}

@article{PhysRevB.80.241101,
  title = {{Fermi} surface of $\alpha$-uranium at ambient pressure},
  author = {Graf, D. and Stillwell, R. and Murphy, T. P. and Park, J.-H. and Kano, M. and Palm, E. C. and Schlottmann, P. and Bourg, J. and Collar, K. N. and Cooley, J. C. and Lashley, J. C. and Willit, J. and Tozer, S. W.},
  journal = {Phys. Rev. B},
  volume = {80},
  issue = {24},
  pages = {241101},
  numpages = {4},
  year = {2009},
  month = {Dec},
  publisher = {American Physical Society},
  doi = {10.1103/PhysRevB.80.241101}
}

@article{PhysRevLett.107.136401,
  title = {Understanding the Complex Phase Diagram of Uranium: The Role of Electron-Phonon Coupling},
  author = {Raymond, S. and Bouchet, J. and Lander, G. H. and Le Tacon, M. and Garbarino, G. and Hoesch, M. and Rueff, J.-P. and Krisch, M. and Lashley, J. C. and Schulze, R. K. and Albers, R. C.},
  journal = {Phys. Rev. Lett.},
  volume = {107},
  issue = {13},
  pages = {136401},
  numpages = {4},
  year = {2011},
  month = {Sep},
  publisher = {American Physical Society},
  doi = {10.1103/PhysRevLett.107.136401},
}

@article{PhysRevLett.126.096401,
  title = {{Quasi-One-Dimensional Fermi Surface Nesting and Hidden Nesting Enable Multiple Kohn Anomalies in $\alpha$-Uranium}},
  author = {Roy, Aditya Prasad and Bajaj, Naini and Mittal, Ranjan and Babu, Peram D. and Bansal, Dipanshu},
  journal = {Phys. Rev. Lett.},
  volume = {126},
  issue = {9},
  pages = {096401},
  numpages = {7},
  year = {2021},
  month = {Mar},
  publisher = {American Physical Society}
}

@article{Lander01021994,
author = {G.H. Lander, E.S. Fisher and S.D. Bader},
title = {The solid-state properties of uranium A historical perspective and review},
journal = {Advances in Physics},
volume = {43},
number = {1},
pages = {1--111},
year = {1994},
publisher = {Taylor \& Francis},
doi = {10.1080/00018739400101465}
}

@article{Dai2014,
  title = {Microscopic evidence for strong periodic lattice distortion in two-dimensional charge-density wave systems},
  author = {Dai, Jixia and Calleja, Eduardo and Alldredge, Jacob and Zhu, Xiangde and Li, Lijun and Lu, Wenjian and Sun, Yuping and Wolf, Thomas and Berger, Helmuth and McElroy, Kyle},
  journal = {Phys. Rev. B},
  volume = {89},
  issue = {16},
  pages = {165140},
  numpages = {5},
  year = {2014},
  month = {Apr},
  publisher = {American Physical Society},
  doi = {10.1103/PhysRevB.89.165140}
}

@article{gu2025,
      title={Pair Wavefunction Symmetry in {UTe$_2$} from Zero-Energy Surface State Visualization}, 
      author={Qiangqiang Gu and Shuqiu Wang and Joseph P. Carroll and Kuanysh Zhussupbekov and Christopher Broyles and Sheng Ran and Nicholas P. Butch and Shanta Saha and Johnpierre Paglione and Xiaolong Liu and J. C. Séamus Davis and Dung-Hai Lee},
journal = {Science},
volume = {388},
number = {6750},
pages = {938-944},
year = {2025},
doi = {10.1126/science.adk7219}
}

@article{Pouget_2024,
doi = {10.1088/1361-6633/ad124f},
year = {2024},
month = {jan},
publisher = {IOP Publishing},
volume = {87},
number = {2},
pages = {026501},
author = {Pouget, Jean-Paul and Canadell, Enric},
title = {Structural approach to charge density waves in low-dimensional systems: electronic instability and chemical bonding},
journal = {Reports on Progress in Physics}
}

@article{pssb.2220560118,
author = {Eschrig, H.},
title = {Unified Perturbation Treatment for Phonons in Metallic Covalent and Ionic Crystals},
journal = {physica status solidi (b)},
volume = {56},
number = {1},
pages = {197-209},
doi = {https://doi.org/10.1002/pssb.2220560118},
year = {1973}
}

@article{PhysRevB.16.3127,
  title = {Superconductivity and phonon softening. {III.} {Relation} between electron bands and phonons in {Nb}, {Mo}, and their alloys},
  author = {Pickett, W. E. and Allen, P. B.},
  journal = {Phys. Rev. B},
  volume = {16},
  issue = {7},
  pages = {3127--3132},
  numpages = {0},
  year = {1977},
  month = {Oct},
  publisher = {American Physical Society},
  doi = {10.1103/PhysRevB.16.3127}
}

@article{smoluchowski1941,
	title = {{Anisotropy of the Electronic Work Function of Metals}},
	volume = {60},
	issn = {0031-899X},
	doi = {10.1103/PhysRev.60.661},
	pages = {661--674},
	number = {9},
	journal = {Phys. Rev.},
	author = {Smoluchowski, R.},
	date = {1941},
	langid = {english},
}

@article{voronoi1908,
  title = {{Nouvelles applications des param{\`e}tres continus {\`a} la th{\'e}orie des formes quadratiques. Deuxi{\`e}me m{\'e}moire. Recherches sur les parall{\'e}llo{\`e}dres primitifs.}},
  author = {Voronoi, Georges},
  year = {1908},
  month = jul,
  journal = {Journal f{\"u}r die reine und angewandte Mathematik (Crelles Journal)},
  volume = {1908},
  number = {134},
  pages = {198--287},
  publisher = {De Gruyter},
  issn = {1435-5345},
  doi = {10.1515/crll.1908.134.198},
  urldate = {2025-10-10},
  chapter = {Journal f{\"u}r die reine und angewandte Mathematik},
  langid = {ngerman},
}

@article{finnis1974,
  title = {Theory of Lattice Contraction at Aluminium Surfaces},
  author = {Finnis, M W and Heine, V},
  year = {1974},
  month = mar,
  journal = {Journal of Physics F: Metal Physics},
  volume = {4},
  number = {3},
  pages = {L37-L41},
  issn = {0305-4608},
  doi = {10.1088/0305-4608/4/3/002},
  urldate = {2025-09-15},
}

@article{bohm1996,
	title        = {Voronoi {{Polyhedra}}: {{A Useful Tool}} to {{Determine}} the {{Symmetry}} and {{Bravais Class}} of {{Crystal Lattices}}},
	shorttitle   = {Voronoi {{Polyhedra}}},
	author       = {Bohm, J. and Heimann, R. B. and Bohm, M.},
	year         = 1996,
	month        = jan,
	journal      = {Crystal Research and Technology},
	volume       = 31,
	number       = 8,
	pages        = {1069--1075},
	doi          = {10.1002/crat.2170310816},
	issn         = {0232-1300, 1521-4079},
	copyright    = {http://onlinelibrary.wiley.com/termsAndConditions\#vor}
}

@article{methfessel1992,
	title        = {Trends of the Surface Relaxations, Surface Energies, and Work Functions of the 4 {\emph{d}} Transition Metals},
	author       = {Methfessel, M. and Hennig, D. and Scheffler, M.},
	year         = 1992,
	month        = aug,
	journal      = {Physical Review B},
	volume       = 46,
	number       = 8,
	pages        = {4816--4829},
	doi          = {10.1103/PhysRevB.46.4816},
	issn         = {0163-1829, 1095-3795},
	copyright    = {http://link.aps.org/licenses/aps-default-license}
}

@article{sun2004,
	title        = {Rule for Structures of Open Metal Surfaces},
	author       = {Sun, Y. Y.},
	year         = 2004,
	journal      = {Physical Review Letters},
	volume       = 93,
	number       = 13,
	doi          = {10.1103/PhysRevLett.93.136102}
}

@article{barnett1983,
  title = {Multilayer Lattice Relaxation at Metal Surfaces: {{A}} Total-Energy Minimization},
  shorttitle = {Multilayer Lattice Relaxation at Metal Surfaces},
  author = {Barnett, R. N. and Landman, Uzi and Cleveland, C. L.},
  year = {1983},
  month = aug,
  journal = {Physical Review B},
  volume = {28},
  number = {4},
  pages = {1685--1695},
  issn = {0163-1829},
  doi = {10.1103/PhysRevB.28.1685},
}

@article{Noonan88,
  title = {Atomic arrangements at metal surfaces},
  author = {Noonan, J.R. and Davis H.L.},
  year = {1988},
  journal = {Science},
  volume = {234},
  pages = {310},
  doi = {10.1126/science.234.4774.310},
}

@misc{bisset2025determiningsuperconductingorderparameter,
      title={Determining the superconducting order parameter of {UPt$_3$} using scanning tunneling microscopy}, 
      author={Rebecca Bisset and Luke C. Rhodes and Hugo Decitre and Matthew J. Neat and Ana Maldonado and Andrew Huxley and Carolina A. Marques and Peter Wahl},
      year={2025},
      volume={arxiv.org/abs/2512.15845}
}

@article{Schmidt10,
  title = {Imaging the Fano lattice to ‘hidden order’ transition in {URu$_2$Si$_2$}},
  author = {A. R. Schmidt and M. H. Hamidian and P. Wahl and F. Meier and A. V. Balatsky and J. D. Garrett and T. J. Williams and G. M. Luke and J. C. Davis },
  year = {2010},
  journal = {Nature},
  volume = {465},
  pages = {570},
  doi = {10.1038/nature09073},
}

@article{Aynajian10,
  title = {Visualizing the formation of the {Kondo} lattice and the hidden order in {URu$_2$Si$_2$}},
  author = {Pegor Aynajian and Eduardo H. da Silva Neto and Colin V. Parker and Yingkai Huang and Abhay Pasupathy and John Mydosh and Ali Yazdani},
  year = {2010},
  journal = {PNAS},
  volume = {107},
  pages = {10383},
  doi = {10.1073/pnas.1005892107},
}

@misc{li2026magneticfieldtunablecommensuratemultiqcharge,
      title={Magnetic-field-tunable commensurate multi-q charge orders on {UTe$_2$} (011) surface}, 
      author={Yuanji Li and Ruotong Yin and Jiashuo Gong and Dengpeng Yuan and Yuguang Wang and Shiyuan Wang and Mingzhe Li and Jiakang Zhang and Ziwei Xue and Zengyi Du and Shiyong Tan and Dong-Lai Feng and Ya-Jun Yan},
      year={2026},
      note={arXiv: 2603.27211}
}

@misc{KondoHybWave,
      title={Observation of {Kondo} hybridization wave in {UTe$_2$}}, 
      author={Xin Yu and Shuikang Yu and Zheyu Wu and Alexander G. Eaton and Andrej Cabala and Michal Vališka and Jun Li and Rui Zhou and Yi-feng Yang and Zhenyu Wang and Peijie Sun and Rui Wu},
      year={2026},
      note={arXiv: 2603.10552}
}

@article{pttl-8m71,
  title = {Quasiparticle Interference of Spin-Triplet Superconductors: Application to {UTe$_2$}},
  author = {Christiansen, Hans and Andersen, Brian M. and Hirschfeld, P. J. and Kreisel, Andreas},
  journal = {Phys. Rev. Lett.},
  volume = {135},
  issue = {21},
  pages = {216001},
  numpages = {7},
  year = {2025},
  month = {Nov},
  publisher = {American Physical Society},
  doi = {10.1103/pttl-8m71}
}

@article{3kcp-w4ny,
  title = {Quasiparticle interference and spectral function of the {UTe$_2$} superconductive surface band},
  author = {Cr\'epieux, Adeline and Pangburn, Emile and Wang, Shuqiu and Zhussupbekov, Kuanysh and Carroll, Joseph P. and Hu, Bin and Gu, Qiangqiang and Davis, J. C. S\'eamus and P\'epin, Catherine and Bena, Cristina},
  journal = {Phys. Rev. B},
  volume = {112},
  issue = {21},
  pages = {214509},
  numpages = {24},
  year = {2025},
  month = {Dec},
  publisher = {American Physical Society},
  doi = {10.1103/3kcp-w4ny}
}

@article{Wu2024,
  title = {Magnetic Signatures of Pressure-Induced Multicomponent Superconductivity in ${\mathrm{UTe}}_{2}$},
  author = {Wu, Zheyu and Chen, Jiasheng and Weinberger, Theodore I. and Cabala, Andrej and Sechovsk\'y, Vladim\'{\i}r and Vali\ifmmode \check{s}\else \v{s}\fi{}ka, Michal and Alireza, Patricia L. and Eaton, Alexander G. and Grosche, F. Malte},
  journal = {Phys. Rev. Lett.},
  volume = {134},
  issue = {23},
  pages = {236501},
  numpages = {7},
  year = {2025},
  month = {Jun},
  doi = {10.1103/PhysRevLett.134.236501}
}

@article{Morr17,
doi = {10.1088/0034-4885/80/1/014502},
year = {2016},
volume = {80},
number = {1},
pages = {014502},
author = {Morr, Dirk K},
title = {Theory of scanning tunneling spectroscopy: from {Kondo} impurities to heavy fermion materials},
journal = {Reports on Progress in Physics}
}

@article{
Hamidian11,
author = {Mohammad H. Hamidian  and Andrew R. Schmidt  and Inês A. Firmo  and Milan P. Allan  and Phelim Bradley  and Jim D. Garrett  and Travis J. Williams  and Graeme M. Luke  and Yonatan Dubi  and Alexander V. Balatsky  and J. C. Davis },
title = {How Kondo-holes create intense nanoscale heavy-fermion hybridization disorder},
journal = {Proceedings of the National Academy of Sciences},
volume = {108},
pages = {18233},
year = {2011},
doi = {10.1073/pnas.1115027108}
}

@article{Dubi2011,
  title = {Hybridization Wave as the ``Hidden Order'' in ${\mathrm{URu}}_{2}{\mathrm{Si}}_{2}$},
  author = {Dubi, Yonatan and Balatsky, Alexander V.},
  journal = {Phys. Rev. Lett.},
  volume = {106},
  pages = {086401},
  numpages = {4},
  year = {2011},
  doi = {10.1103/PhysRevLett.106.086401}
}

@article{Hafner22,
  title = {Possible Quadrupole Density Wave in the Superconducting {Kondo} Lattice ${\mathrm{CeRh}}_{2}{\mathrm{As}}_{2}$},
  author = {Hafner, D. and Khanenko, P. and Eljaouhari, E.-O. and K\"uchler, R. and Banda, J. and Bannor, N. and L\"uhmann, T. and Landaeta, J. F. and Mishra, S. and Sheikin, I. and Hassinger, E. and Khim, S. and Geibel, C. and Zwicknagl, G. and Brando, M.},
  journal = {Phys. Rev. X},
  volume = {12},
  issue = {1},
  pages = {011023},
  numpages = {10},
  year = {2022},
  month = {Feb},
  publisher = {American Physical Society},
  doi = {10.1103/PhysRevX.12.011023}
}

@article{Huang19,
  title = {Charge density wave in a doped {Kondo} chain},
  author = {Huang, Yixuan and Sheng, D. N. and Ting, C. S.},
  journal = {Phys. Rev. B},
  volume = {99},
  pages = {195109},
  year = {2019},
  doi = {10.1103/PhysRevB.99.195109}
}

@article{Wang25,
  title = {Odd-parity quasiparticle interference in the superconductive surface state of {UTe$_2$}},
  author = {Shuqiu Wang and Kuanysh Zhussupbekov and Joseph P. Carroll and Bin Hu and Xiaolong Liu and Emile Pangburn and Adeline Crepieux and Catherine Pepin and Christopher Broyles and Sheng Ran and Nicholas P. Butch and Shanta Saha and Johnpierre Paglione and Cristina Bena and J. C. Séamus Davis and Qiangqiang Gu },
  journal = {Nat. Phys.},
  volume = {21},
  pages = {1555},
  year = {2025},
  doi = {10.1038/s41567-025-03000-w}
}

@article{Yin25,
  title = {Yin-Yang vortex on {UTe$_2$} (011) surface},
  author = {Ruotong Yin and Yuanji Li and Zengyi Du and Dengpeng Yuan and Shiyuan Wang and Jiashuo Gong and Mingzhe Li and Ziyuan Chen and Jiakang Zhang and Yuguang Wang and Ziwei Xue and Xinchun Lai and Shiyong Tan and Da Wang and Qiang-Hua Wang and Dong-Lai Feng and Ya-Jun Yan},
  journal = {Nat. Comms.},
  volume = {17},
  pages = {5394},
  year = {2026},
  doi = {10.1038/s41467-026-72162-9}
}

@article{Sharma25,
  title = {Observation of Persistent Zero Modes and Superconducting Vortex Doublets in {UTe$_2$}},
  author = {Nileema Sharma and Matthew Toole and James McKenzie and Fangjun Cheng and Mitchell M. Bordelon and Sean M. Thomas and Priscila F. S. Rosa and Yi-Ting Hsu and Xiaolong Liu},
  journal = {ACS Nano},
  volume = {19},
  pages = {31539},
  year = {2025},
  doi = {10.1021/acsnano.5c08406}
}

@article{Yang25,
  title = {Spectroscopic evidence of symmetry breaking in the superconducting vortices of {UTe$_2$}},
  author = {Zhongzheng Yang and Fanbang Zheng and Dingsong Wu and Bin-Bin Zhang and Ning Li and Wenhui Li and Chaofan Zhang and Guang-Ming Zhang and Xi Chen and Yulin Chen and Shichao Yan},
  journal = {National Science Review},
  volume = {12},
  pages = {nwaf267},
  year = {2025},
  doi = {10.1093/nsr/nwaf267}
}

@article{Wolfle10,
  title = {Tunneling into Clean Heavy Fermion Compounds: Origin of the {Fano} Line Shape},
  author = {W\"olfle, P. and Dubi, Y. and Balatsky, A. V.},
  journal = {Phys. Rev. Lett.},
  volume = {105},
  pages = {246401},
  numpages = {4},
  year = {2010},
  doi = {10.1103/PhysRevLett.105.246401}
}

@article{Figgins10,
  title = {Differential Conductance and Quantum Interference in {Kondo} Systems},
  author = {Figgins, Jeremy and Morr, Dirk K.},
  journal = {Phys. Rev. Lett.},
  volume = {104},
  pages = {187202},
  year = {2010},
  doi = {10.1103/PhysRevLett.104.187202}
}

@article{Maltseva09,
  title = {Electron Cotunneling into a {Kondo} Lattice},
  author = {Maltseva, Marianna and Dzero, M. and Coleman, P.},
  journal = {Phys. Rev. Lett.},
  volume = {103},
  pages = {206402},
  year = {2009},
  doi = {10.1103/PhysRevLett.103.206402}
}
\end{document}